\documentclass[reprint,pra,twocolumn,showpacs,superscriptaddress,aps,longbibliography]{revtex4-2}

\usepackage[colorlinks]{hyperref}
\hypersetup{%
        plainpages=true,
        breaklinks=true,
        hypertexnames=false,
        pageanchor=true,
        colorlinks=true,
        linkcolor={blue},
        citecolor={blue},
        urlcolor={blue},
        anchorcolor={black}
      }

\usepackage{graphicx}
\usepackage{bm}
\usepackage{amsmath}
\usepackage{amssymb}

\DeclareGraphicsExtensions{.png,.jpg,.eps}
\usepackage{xcolor}

\newcommand{\beq}{\begin{equation}}
\newcommand{\eeq}{\end{equation}}

\usepackage{mleftright} 

\newcommand{\figref}[1]{\mbox{Fig.~\ref{#1}}}

\newcommand{\secref}[1]{\mbox{Section~\ref{#1}}}

\newcommand{\appref}[1]{\mbox{Appendix~\ref{#1}}}
\renewcommand{\eqref}[1]{\mbox{Eq.~(\ref{#1})}}

\newcommand{\figpanel}[2]{Fig.~\hyperref[#1]{\ref*{#1}(#2)}}
\newcommand{\figpanels}[3]{Fig.~\hyperref[#1]{\ref*{#1}(#2)--(#3)}}
\newcommand{\figpanelNoPrefix}[2]{\hyperref[#1]{\ref*{#1}(#2)}}

\newcommand{\ket}[1]{|#1 \rangle}
\newcommand{\bra}[1]{\langle #1|}

\AtBeginDocument{
\renewcommand{\tilde}[1]{#1'}
}

\usepackage{siunitx}

\begin{document}




\author{Guangze Chen}
\affiliation{Department of Microtechnology and Nanoscience, Chalmers University of Technology, 41296 Gothenburg, Sweden}

\author{Anton Frisk Kockum}
\affiliation{Department of Microtechnology and Nanoscience, Chalmers University of Technology, 41296 Gothenburg, Sweden}

\title{Simulating open quantum systems with giant atoms}

\begin{abstract}
Open quantum many-body systems are of both fundamental and applicational interest. However, it remains an open challenge to simulate and solve such systems, both with state-of-the-art classical methods and with quantum-simulation protocols. To overcome this challenge, we introduce a simulator for open quantum many-body systems based on giant atoms, i.e., atoms (possibly artificial), that couple to a waveguide at multiple points, which can be wavelengths apart. We first show that a simulator consisting of two giant atoms can simulate the dynamics of two coupled qubits, where one qubit is subject to different drive amplitudes and dissipation rates. This simulation enables characterizing the quantum Zeno crossover in this model. We further show that by equipping the simulator with post-selection, it becomes possible to simulate the effective non-Hermitian Hamiltonian dynamics of the system and thereby characterize the transition from oscillatory to non-oscillatory dynamics due to varying dissipation rates. We demonstrate and analyze the robustness of these simulation results against noise affecting the giant atoms. Finally, we discuss and show how giant-atom-based simulators can be scaled up for digital-analog simulation of large open quantum many-body systems, e.g., generic dissipative spin models.
\end{abstract}

\date{\today}

\maketitle


\section{Introduction}

Open quantum systems~\cite{breuer2002theory} have attracted much research interest for a long time. Unlike their closed counterparts with purely coherent dynamics, these systems also display dissipative dynamics resulting from coupling to surrounding environments. Such coupling is inevitable to some degree in realistic physical systems; therefore, open quantum systems are important for describing realistic setups in quantum optics, quantum chemistry, and materials science~\cite{McArdle2020, Hbener2020, Blais2021, sieberer2023universality}. The interplay between coherent and dissipative dynamics in open quantum systems enables the engineering of exotic steady states with designed interaction and dissipation~\cite{Harrington2022, Diehl2008, Kraus2008, Verstraete2009, Diehl2010, Diehl2011, Reiter2016, bardyn2012majorana, Mi2024}. Furthermore, open quantum systems exhibit unique dynamics without a counterpart in closed quantum systems, where many-body physics~\cite{Cai2013, Wybo2020, Olmos2012, Znidaric2015, Bouganne2020, Wang2020Hierarchy, Wang2024Embedding, Chen2024} and non-Hermitian topology~\cite{Song2019, Haga2021, Okuma2021} can be involved to result in intriguing phenomena.


Despite the intense interest in open quantum systems, it remains an open challenge to simulate and solve such systems when many-body interactions are present. For classical simulation methods~\cite{Weimer2021, Luchnikov2019, Aloisio2023}, there are two main parts to this challenge: (i) the quantum many-body nature of the system makes the simulation complexity scale exponentially with the system size and (ii) the openness means that a more extensive description of the system state is required compared to a closed system. Quantum simulation~\cite{Feynman1982, Lloyd1996, Georgescu2014, Cirac2012_Goals}, where one quantum system is used to simulate another, addresses both parts of the challenge and therefore offers the possibility to investigate open quantum many-body systems beyond the capability of classical methods~\cite{trivedi2023quantum, Delgado-Granados2024}. However, quantum simulation of generic open quantum many-body systems requires a simultaneously scalable and highly tunable simulator, which is not yet available. For example, scalable simulators such as purely analog simulators using cold atoms~\cite{Bloch2012, Gross2017, Schafer2020} or trapped ions~\cite{Barreiro2011, Schindler2013, Schlawin2021, Chertkov2023, Kang2024, Qiao2024, zhang2024observation, sun2024quantum} are restricted to the intrinsic physical models in these systems. At the same time, simulators with greater tunability, able to simulate a larger variety of models, often require a more complicated physical setup. In particular, digital simulators~\cite{Fauseweh2024} of open quantum systems, e.g., using superconducting qubits, often require ancillary qubits to mimic the environment~\cite{Han2021, Metcalf2022, Mi2024}. Furthermore, tunable qubit-qubit and qubit-environment couplings usually require additional parametric (or otherwise tunable) couplers~\cite{Mostame2012, McKay2016, Bengtsson2020, Kannan2023, Yang2023}. This complexity of the physical setups impairs the scaling of such quantum simulators to larger system sizes.

\begin{figure}
\center
\includegraphics[width=\linewidth]{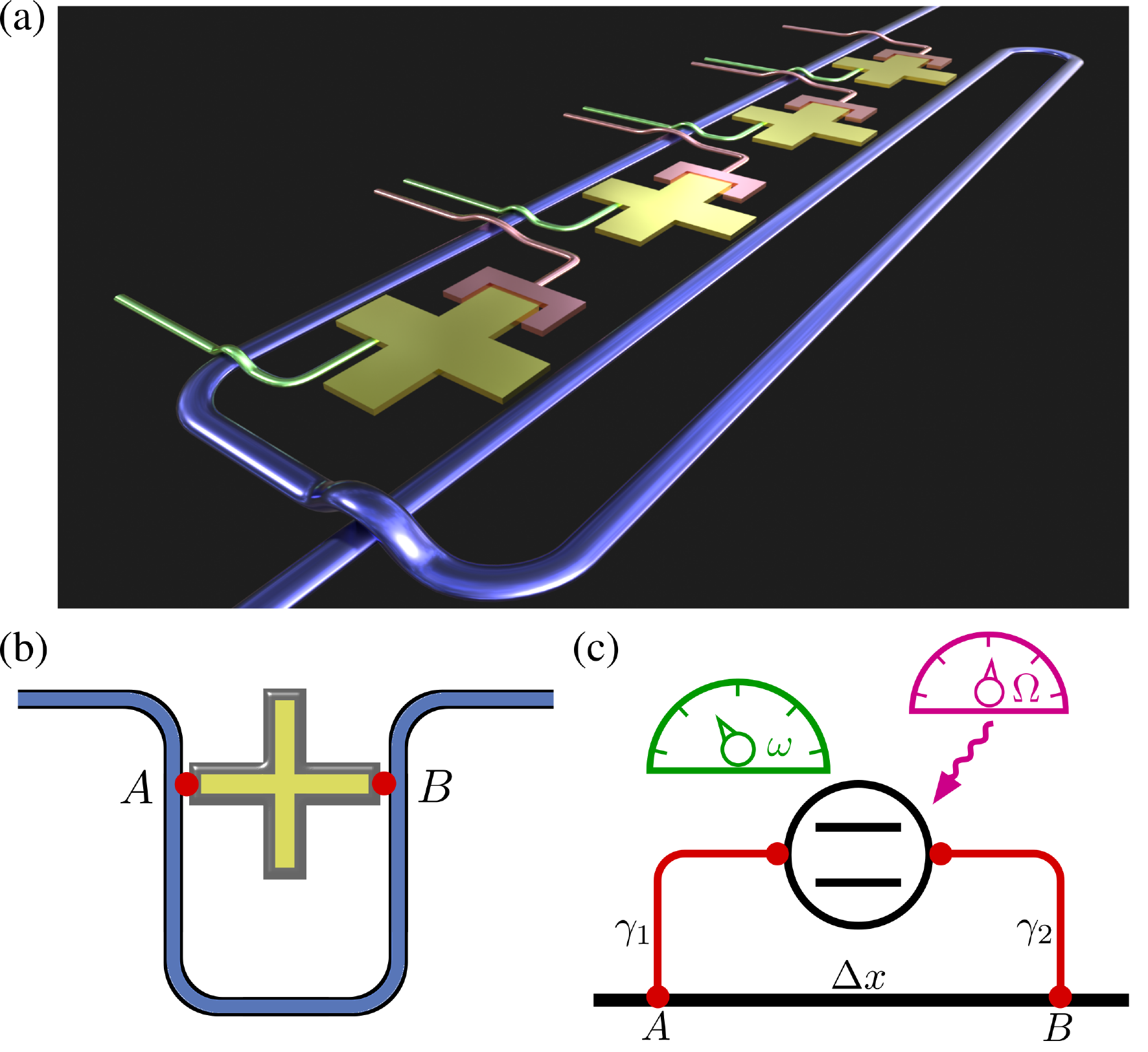}
\caption{A quantum simulator based on giant atoms and its basic building blocks.
(a) An artistic rendition of a four-qubit version of the quantum simulator, where superconducting transmon qubits (yellow) are coupled to a waveguide (blue). A flux line (green; used to control the qubit frequency), and a readout resonator (pink) are coupled to each qubit. 
(b) Each qubit in panel (a) couples capacitively to the waveguide at two points $A$ and $B$ (red dots), which are separated by a distance on the order of the wavelength of the light propagating in the waveguide. This makes the qubit a giant (artificial) atom.
(c) A sketch of a giant atom and its tunable parameters. The coupling strengths to the waveguide ($\gamma_1$ and $\gamma_2$ at points $A$ and $B$, respectively) and the distance $\Delta x$ between the coupling points are generally fixed in fabrication. However, the flux line enables tuning the qubit transition frequency $\omega$, and an external drive of strength $\Omega$ can be applied to the qubit through the resonator. The frequency $\omega$ will in turn set the effective relaxation rate $\Gamma$ of the qubit to the environment (the waveguide), as well as the strength $g$ of its interaction with other qubits through the waveguide.}
\label{fig0}
\end{figure}

To circumvent the drawbacks of existing quantum-simulation setups for open quantum systems, we here introduce a scalable and highly tunable quantum simulator based on giant artificial atoms~\cite{Kockum2021}; see \figref{fig0}. While a traditional small (artificial) atom can be approximated as point-like when comparing its size to the wavelength of the light it interacts with, a giant atom couples to its surroundings at multiple discrete points, which can be wavelengths apart, as illustrated in \figpanels{fig0}{b}{c}. Interference effects due to having these multiple coupling points endow giant atoms with frequency-dependent relaxation rates~\cite{Kockum2014, Vadiraj2021} and qubit-qubit interaction strengths~\cite{Kockum2018}; tuning the frequency of a giant atom thus enables tuning several other system parameters across a wide range of values, which is important for quantum simulation. For example, two-qubit gates have been performed on giant atoms in the form of superconducting qubits without additional couplers, just by tuning the frequencies of the artificial atoms~\cite{Kannan2020}. In addition to these capabilities, other fundamental properties of giant atoms have been investigated intensively in the past few years, both in theory~\cite{Guo2017, Gonzalez-Tudela2019, Guo2020, Guimond2020, Gheeraert2020, Ask2020, Wang2021, Du2021, Soro2022, Wang2022, Du2022, Du2022a, Terradas-Brianso2022, Soro2023, Du2023, Ingelsten2024, Wang2024} and in experiments~\cite{Gustafsson2014, Manenti2017, Satzinger2018, Moores2018, Sletten2019, Bienfait2019, Andersson2019, Bienfait2020, Andersson2020, Wang2022a, Kannan2023, Joshi2023}. This well-developed theoretical understanding and experimental realization of giant atoms have prepared them for applications in quantum simulation and other quantum technologies.

We demonstrate how a giant-atom-based quantum simulator works by starting from an example of two giant atoms that simulate two coupled qubits, where one of the qubits is subject to both dissipation and a coherent drive. In particular, we show that, by tuning the frequency of one of the giant atoms, our simulator can simulate the Liouvillian dynamics of such a model at different dissipation rates and drive strengths, which enables us to characterize the quantum Zeno crossover~\cite{Li2023, Dai2023} in this model. We further show that by performing post-selection~\cite{Li2018, Naghiloo2019, Skinner2019} in the giant-atom simulator, we can simulate the effective non-Hermitian Hamiltonian dynamics of the two-qubit model. In particular, the simulator with post-selection can characterize a transition from oscillatory to non-oscillatory dynamics in this model that occurs when varying the ratio between the drive strength and the dissipation rate. We discuss and quantify the robustness of all these simulation results against various possible imperfections in the quantum simulator, such as relaxation or dephasing of the giant atoms due to interaction with some other environment than the waveguide.

Moving beyond the two-qubit example, we next show how to arrange many giant atoms in scalable simulators capable of handling generic dissipative quantum spin systems. The key to this capability is that single-qubit gates implemented by driving the giant atoms and two-qubit gates performed by tuning the frequencies of the giant atoms together form a universal gate set, which can simulate any Hamiltonian dynamics. The ability to change the coupling to the waveguide by changing the giant-atom frequencies extends the capability of the simulator to include dissipation. We show that such giant-atom simulators have better scalability than conventional small-atom simulators due to simpler structures and fewer required atoms. We also present how our giant-atom simulators can be realized in experiments with superconducting circuits and discuss possible scaling limitations, e.g., when distances between giant atoms or their coupling points lead to non-Markovian effects.

The rest of this article is organized as follows. In \secref{sec_sim_idea}, we outline the basic theory of the giant-atom quantum simulator, showing how a Trotter--Suzuki decomposition of Liouvillian open-system dynamics can be implemented by tuning the frequencies of giant atoms. We then move to a specific illustrative example: in \secref{Ch_model}, we present the model of a qubit coupled to a driven-dissipative qubit, and show its Liouvillian and effective Hamiltonian dynamics. In \secref{Ch_sim}, we show how a simulator consisting of two giant atoms can implement a quantum simulation of this model. In particular, we show that the giant-atom simulator can faithfully capture the quantum Zeno crossover in the Liouvillian dynamics and the transition from oscillatory to non-oscillatory dynamics in the effective non-Hermitian Hamiltonian dynamics of the model. We then analyze, in \secref{Ch_err}, the robustness of the simulation results against errors due to finite qubit lifetimes and dephasing times at the levels seen in state-of-the-art experimental platforms. In \secref{Ch_scale_up}, we show how giant-atom-based quantum simulators can be scaled to more giant atoms and that they enable simulation of generic dissipative spin systems. We conclude in \secref{Ch_conclusion} with a summary of our results and an outlook. A few details and derivations are relegated to appendixes: \appref{app_L} gives further information about the two-qubit model used in our illustrative example, \appref{app_H} derives non-Hermitian Hamiltonians resulting from post-selection, \appref{app_omega} gives further details about how to tune the frequencies of giant atoms in our simulator, and \appref{sec_app_err} discusses a few additional potential error sources for the simulator.


\section{General idea for quantum simulation with giant atoms}
\label{sec_sim_idea}

Here we present the idea behind using giant atoms for quantum simulation of open quantum systems. We first review how the time evolution of an open quantum system can be decomposed into sequences of short time steps that each just implements some part of the coherent or dissipative dynamics for the system. We then explain how two giant atoms coupled to a waveguide constitute a fundamental quantum-simulation unit that can be controlled to realize all such steps.

The time evolution of a Markovian open quantum system is given by a Lindblad master equation~\cite{Lindblad1976, Gorini1976, breuer2002theory} ($\hbar = 1$ throughout this article)
\beq
\partial_t \rho = - i \mleft[ H, \rho \mright] + \sum_{k} \mathcal{D}[X_k] \rho ,
\label{eq:Lindblad}
\eeq
where $\rho$ is the density matrix of the system, $H$ is the system Hamiltonian, the $X_k$ are system operators coupling to a surrounding environment, and $\mathcal{D}[X_k] \rho = X_k \rho X_k^\dag - \frac{1}{2} X_k^\dag X_k \rho - \frac{1}{2} \rho X_k^\dag X_k$ are Lindblad operators. This equation can be written more compactly as
\beq
\partial_t \rho = \mathcal{L} \rho ,
\label{eq:LiouvillianME}
\eeq
where $\mathcal{L}$ is the Liouvillian, and has the solution
\beq
\rho(t) = \exp \mleft( \mathcal{L} t \mright) \rho (0),
\eeq
given an initial state $\rho (t = 0)$.

To simulate this time evolution generated by a generic Liouvillian $\mathcal{L}$ not intrinsically present in the simulator, a standard approach is to consider an expansion of it into parts. Writing
\beq
\mathcal{L} = \sum_{j=1}^n \mathcal{L}_j ,
\eeq
where each superoperator $\mathcal{L}_j$ generates parts of the coherent and/or dissipative dynamics in \eqref{eq:Lindblad}, a first-order Trotter--Suzuki decomposition~\cite{SUZUKI1990319, Kliesch2011} of the time-evolution operator becomes
\beq
\exp \mleft( \mathcal{L} t \mright) = \mleft[ \prod_{j=1}^n \exp\mleft( \mathcal{L}_j t/l \mright) \mright]^l + O \mleft( \frac{t^2}{l} \mright) .
\label{eq:TrotterSuzuki}
\eeq
Given that we divide the time evolution into enough steps $l$ that the error becomes negligible, the task of simulating a many-body Liouvillian $\mathcal{L}$ is thus reduced to simulating simpler components $\mathcal{L}_j$ acting on few-body subspaces (assuming that the Liouvillian is local). Notably, for an open quantum system, this decomposition enables us to separate the dissipative and coherent dynamics in $\mathcal{L}$. By adjusting the lengths of the time steps associated with each part, we can thus change their relative strengths and study the competition between these components in the dynamics, which is one of the main directions of the study of open quantum systems~\cite{sieberer2023universality}.

The major challenge for implementing a quantum simulator relying on \eqref{eq:TrotterSuzuki} is being able to turn the different components $\mathcal{L}_j$ on and off without too much overhead in resources such as ancillary qubits or complicated tunable coupling elements between qubits and some environment. The essential property of a giant-atom quantum simulator is that it overcomes this challenge by being able to turn on and off coherent and dissipative dynamics for its components solely by tuning the frequencies of its qubits, without the need for extra coupling elements.

\begin{figure}
\center
\includegraphics[width=\linewidth]{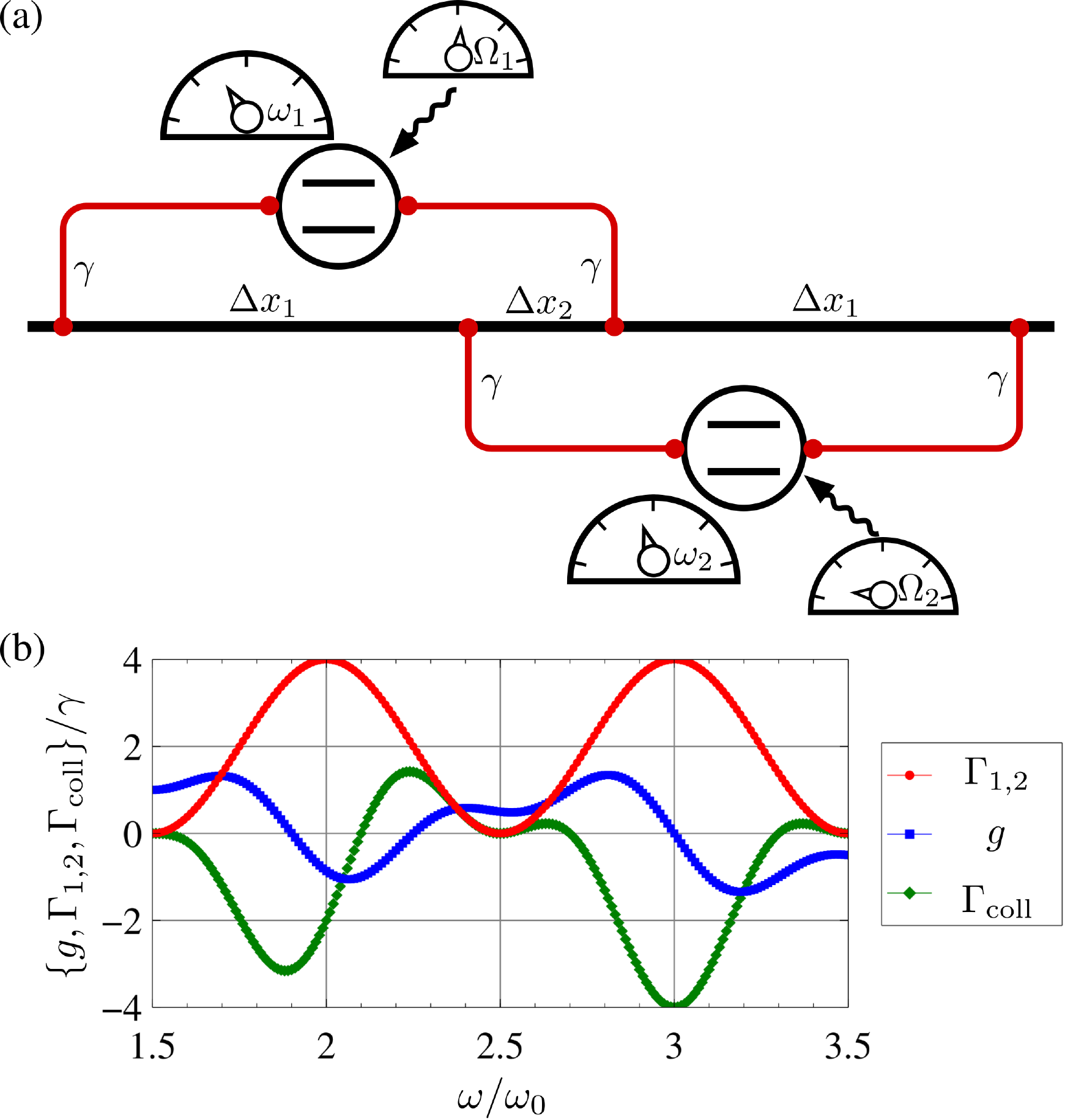}
\caption{The essential properties of a giant-atom quantum simulator.
(a) A fundamental unit in such a simulator, consisting of two giant atoms with tunable frequencies $\omega_k$ and drives $\Omega_k$. The giant atoms (qubits) are coupled to the waveguide at two points each; every such coupling has strength $\gamma$.
(b) The waveguide-mediated coherent coupling $g$ between qubits, individual qubit decay rates $\Gamma_k$, and the collective decay $\Gamma_\text{coll}$ of the two qubits as a function of frequency $\omega = \omega_1 = \omega_2$ for $\Delta x_1 = 5 \Delta x_2$. Tuning the qubit frequencies $\omega_k$ changes all these parameters determining coherent and dissipative dynamics, which enables quantum simulation of different parameter regimes.}
\label{fig_simulator}
\end{figure}

This key functionality of giant atoms can be fully explained by considering the setup shown in \figpanel{fig_simulator}{a}. In this setup, two giant atoms are coupled to a waveguide at two points each in a ``braided'' topology, i.e., with one coupling point of each atom located in between the coupling points of the other atom. Tracing out the waveguide degrees of freedom by assuming Markovianity and viewing each atom as a two-level system (a qubit), the master equation for the atomic degrees of freedom is~\cite{Kockum2018}
\begin{align} \label{eq_simulator_L}
\partial_t \rho &= -i \mleft[ \omega_1 \frac{\sigma_1^z}{2} + \omega_2 \frac{\sigma_2^z}{2} + g (\omega_1, \omega_2) \mleft( \sigma_1^+ \sigma_2^- + \text{H.c.} \mright) \mright. \nonumber\\
&\qquad\quad + \Omega_1(t) \sigma_1^x + \Omega_2(t) \sigma_2^x , \rho \bigg] \nonumber \\ 
&\quad + \Gamma_1(\omega_1) \mathcal{D}[\sigma_1^-] \rho + \Gamma_2(\omega_2) \mathcal{D}[\sigma_2^-] \rho \nonumber \\
&\quad + \Gamma_\text{coll}(\omega_1,\omega_2) \mleft[ \mleft( \sigma_1^- \rho \sigma_2^+ - \frac{1}{2} \mleft\{ \sigma_1^+ \sigma_2^- , \rho \mright\} \mright) + \text{H.c.} \mright],
\end{align}
where $\omega_k$ is the transition frequency of qubit $k$, $\sigma_k^z$ ($\sigma_k^x$) is the Pauli $Z$ ($X$) matrix of qubit $k$, $\sigma_k^+$ ($\sigma_k^-$) is the raising (lowering) operator of qubit $k$, $\Omega_k$ is the strength of the coherent drive on qubit $k$, $g$ is the strength of the effective coherent coupling between the qubits mediated by the waveguide, $\Gamma_k$ is the individual decay rate of qubit $k$, $\Gamma_\text{coll}$ is the collective decay rate of the qubits, and H.c.~denotes Hermitian conjugate. 

The interference between emission from different coupling points in the giant atoms makes $g$, $\Gamma_k$, and $\Gamma_\text{coll}$ functions of the qubit frequencies. An example of how these frequency dependencies can look is given in \figpanel{fig_simulator}{b}. There, we have set $\omega_1 = \omega_2 \equiv \omega$, assumed equal coupling strengths $\gamma$ at every coupling point, defined distances $\Delta x_1$ and $\Delta x_2$ between coupling points as shown in \figpanel{fig_simulator}{a} and set $\Delta x_1 = 5 \Delta x_2$, and defined $\omega_0 = 2 \pi v / (\Delta x_1 + \Delta x_2)$ with $v$ the speed of light in the waveguide. In particular, these settings yield $\Gamma_k(\omega_k) = 2 \gamma \mleft[ 1 + \cos \mleft( 2 \pi \omega_k / \omega_0 \mright) \mright]$. We observe that there is a point $\omega_\text{DF} = 2.5 \omega_0$, where $g = 0.5 \gamma$ while both $\Gamma_k = 0$ and $\Gamma_\text{coll} = 0$. This decoherence-free interaction can only occur with braided giant atoms; it is not possible with small atoms or other configurations of giant atoms.

The decoherence-free interaction enables performing a two-qubit XY gate in the system, as demonstrated in an experiment with superconducting qubits~\cite{Kannan2020}. Since we also can drive each qubit coherently with strength $\Omega_k$ and perform virtual Z gates, all while parking the qubits at frequencies where $\Gamma_k = 0$ and $\Gamma_\text{coll} = 0$, we have access to a universal gate set to simulate any coherent dynamics. Furthermore, we can turn off all coherent dynamics and turn on dissipation with a strength of our choice. For example, when $\omega_2 = \omega_\text{DF}$ and $\omega_1 = 2 \omega_0$, we have $\Gamma_1 = 4 \gamma$ and $\Gamma_2 = \Gamma_\text{coll} = 0$, i.e., only decay from qubit 1. Since $g \ll \omega_k$ in the physical setups we consider, the qubit-qubit coupling here is negligible compared to the detuning of $0.5 \omega_0$. In a similar manner, we can achieve other purely dissipative dynamics in the system by changing the frequencies $\omega_k$.

Since all interactions in a setup with many giant atoms are pairwise, the example here with two giant atoms provides the necessary understanding also for larger setups. We have thus shown the capability of a giant-atom quantum simulator to achieve generic coherent and dissipative dynamics separately, meaning that we can implement the method of \eqref{eq:TrotterSuzuki} for quantum simulation of open quantum systems. The details and advantages of such an implementation will depend on the model to be simulated. To provide a concrete example of such details, we present in the following sections the simulation of a particular model, where the competition between coherent and dissipative dynamics results in a quantum Zeno crossover~\cite{Li2023, Dai2023}. 


\section{A model to simulate --- quantum Zeno crossover for two qubits}
\label{Ch_model}

\begin{figure*}
\center
\includegraphics[width=\linewidth]{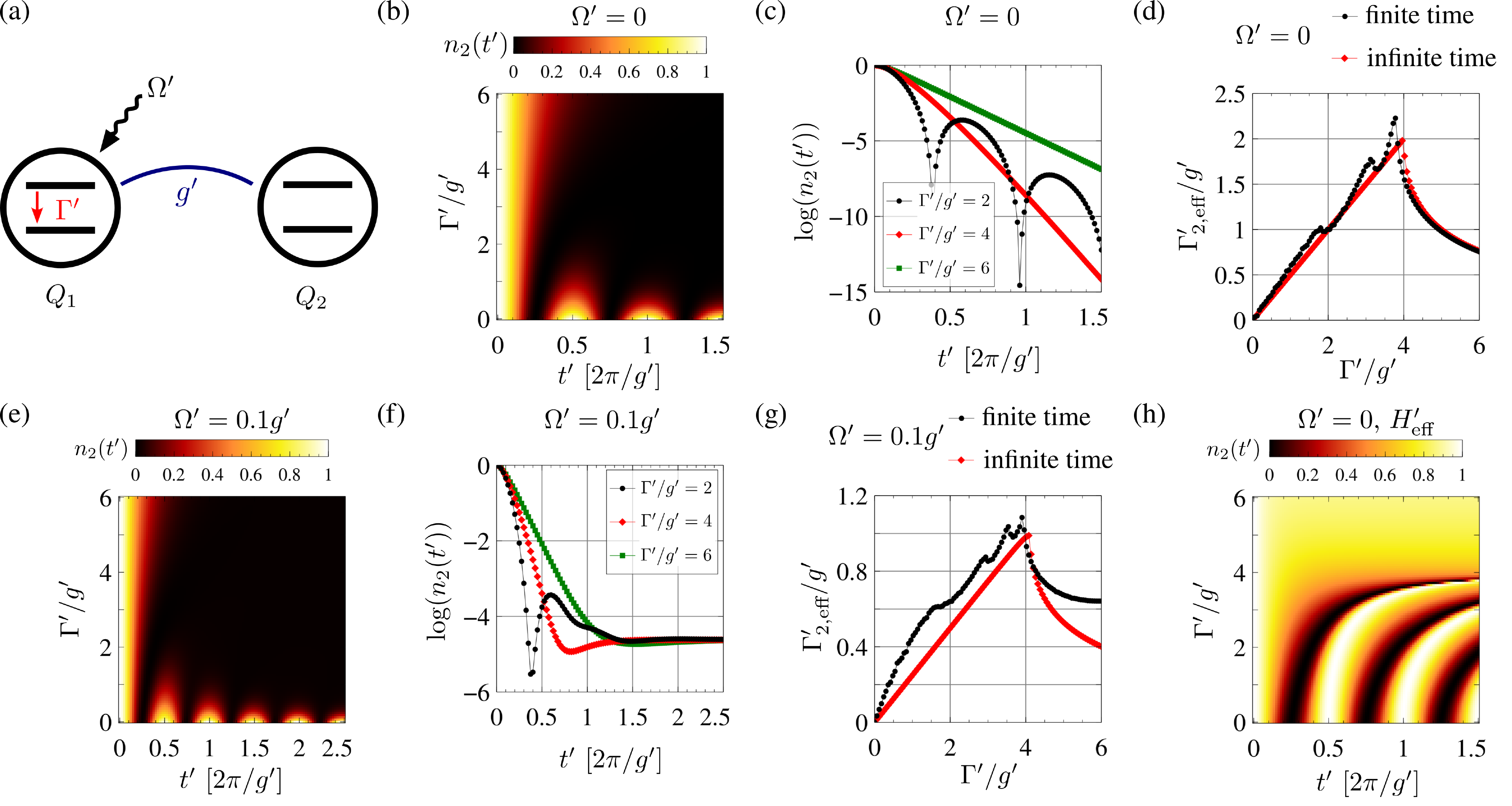}
\caption{The two-qubit model to be simulated and its dynamics.
(a) Sketch of the model governed by \eqref{eq_model_L}, where a first qubit $Q_1$ is subject to both a coherent drive $\Omega'$ and dissipation $\Gamma'$, and is coupled with strength $g$ to a second qubit $Q_2$, which is otherwise isolated from the surroundings.
(b) The time evolution of $n_2(t')$, the population of $Q_2$, for $\Omega' = 0$ and different $\Gamma'$. The initial state is $\rho(0) = \mleft( \ket{0}_1 \otimes \ket{1}_2 \mright) \mleft( \bra{0}_1 \otimes \bra{1}_2 \mright)$, i.e., $n_2(0) = 1$.
(c) Linecuts from panel (b) showing $n_2(t')$ on a logarithmic scale for $\Gamma' = \{2, 4, 6\} g'$.
(d) The effective relaxation rate $\Gamma'_{2, \text{eff}}$ of $Q_2$ as a function of $\Gamma'$ for $\Omega'=0$. The infinite-time dynamics (red) indicates a quantum Zeno crossover at $\Gamma'=4g'$, where $\Gamma'_{2, \text{eff}}$ reaches its maximum. The finite-time dynamics from $t'=0$ to $t' = 3 \pi / g'$ (black) predicts this crossover at $\Gamma' \approx 3.8 g'$. 
(e) The time evolution of $n_2(t')$ for $\Omega' = 0.1 g'$.
(f) Linecuts from panel (e) showing $n_2(t')$ on a logarithmic scale for $\Gamma' = \{2, 4, 6\} g'$.
(g) The effective relaxation rate $\Gamma'_{2, \text{eff}}$ of $Q_2$ as a function of $\Gamma'$ for $\Omega' = 0.1 g'$. Compared to the case $\Omega'=0$ in panel (d), $\Gamma'_{2, \text{eff}}$ is reduced and the quantum Zeno crossover point is shifted to around $\tilde{\Gamma} \approx 4.11 \tilde{g}$ (infinite-time dynamics; red). The finite-time dynamics from $t'=0$ to $t'=5\pi/g'$ (black) predicts this crossover at $\Gamma' \approx 3.9g'$.
(h) The evolution of $n_{2}(t')$ under the effective non-Hermitian Hamiltonian in \eqref{eq_model_Heff} for $\Omega'=0$. The time evolution shows a transition from oscillatory to non-oscillatory behavior at $\Gamma' \approx 3.8g'$.
\label{fig_model}}
\end{figure*}

As our illustrative example for quantum simulation, we take a model of two coupled qubits, where the first 
qubit is subject to both a coherent drive and dissipation, while the second qubit is isolated from its surroundings except for its coupling to the first qubit. This model is sketched in \figpanel{fig_model}{a}. Its dynamics are given by the master equation
\begin{equation} \label{eq_model_L}
\partial_t \rho = \tilde{\mathcal{L}} \rho = -i \mleft[ \tilde{g} \mleft( \sigma_1^+ \sigma_2^- + \text{H.c.} \mright) + \tilde{\Omega} \sigma_1^x , \rho \mright] - \tilde{\Gamma} \mathcal{D}[\sigma_1^-] \rho ,
\end{equation}
where $\tilde{g}$ is the strength of the coupling between the qubits, $\tilde{\Omega}$ is the amplitude of the drive on qubit 1, and $\tilde{\Gamma}$ is the decay rate of qubit 1. The master equation is written in the rotating frame of the qubit frequencies $\omega_1' = \omega_2'$. The prime on the parameters indicates that they are parameters to be simulated, and as such differ from the physical parameters in a simulator, which will be written without any tilde. 

In this paradigmatic model, the competition between the coherent and dissipative dynamics results in a quantum Zeno crossover~\cite{Li2023, Dai2023} at $\tilde{\Gamma} = 4 \tilde{g}$~\cite{Minganti2019} for $\tilde{\Omega} = 0$. At this point, the maximum decay rate for an arbitrary initial state is obtained, which is important, e.g., for quantum state transfer~\cite{Yang2023}. 

To show how the quantum Zeno crossover manifests in this model, we consider an initial state with qubit 1 in its ground state and qubit 2 excited: $\rho (0) = \mleft( \ket{0}_1 \otimes \ket{1}_2 \mright) \mleft( \bra{0}_1 \otimes \bra{1}_2 \mright)$. In \figpanel{fig_model}{b}, we plot the time evolution of the population of qubit 2, $n_2(\tilde{t}) = \mleft\{ 1 + \text{Tr} \mleft( \sigma_2^z \rho(\tilde{t}) \mright] \mright\} / 2$, where $\rho(\tilde{t}) = \exp \mleft(\tilde{\mathcal{L}} \tilde{t} \mright) \rho(0)$. By looking at the logarithm of $n_2(\tilde{t})$ for a few linecuts from \figpanel{fig_model}{b} in \figpanel{fig_model}{c}, we see that the population of qubit 2 decays faster when $\tilde{\Gamma} = 4 \tilde{g}$ than when $\tilde{\Gamma} = 2 \tilde{g}$ or $\tilde{\Gamma} = 6 \tilde{g}$. To obtain the effective relaxation rate $\Gamma'_{2, \text{eff}}$ of qubit 2, we fit $\log \mleft[ n_2(\tilde{t}) \mright]$ to the linear form $-\Gamma'_{2, \text{eff}} \tilde{t}+C$. Plotting the resulting $\Gamma'_{2, \text{eff}}$ in \figpanel{fig_model}{d}, we see that it increases (decreases) with $\Gamma'$ for weak (strong) $\Gamma'$, which is known as the quantum anti-Zeno (Zeno) effect~\cite{Li2023, Catalano2023}. These two regimes are separated by the quantum Zeno crossover point $\tilde{\Gamma} \approx 3.8 \tilde{g}$ where $\Gamma'_{2, \text{eff}}$ reaches its maximum. Note that we here considered finite-time dynamics, since that is what is feasible for quantum simulation. Therefore the predicted quantum Zeno crossover point has an error compared to that obtained from infinite-time dynamics (see \appref{app_L}), which can be reduced by increasing $t'$.

When the external drive is turned on, i.e., $\tilde{\Omega}\neq0$, it results in a change of the steady state of the system; see \figpanel{fig_model}{e,f} for the same plots as in \figpanel{fig_model}{b,c} with $\tilde{\Omega} \neq 0$. In particular, $n_2(\tilde{t}\to\infty)\neq0$ in this case. We therefore fit the relaxation rate $\Gamma'_{2, \text{eff}}$ as $\log(n_2(\tilde{t})-n_2(\tilde{t}_f))\approx-\Gamma'_{2, \text{eff}} \tilde{t}+C$, choosing the final time for the simulation to be $\tilde{t}_f = 5 \pi / \tilde{g}$. Plotting the resulting $\Gamma'_{2, \text{eff}}$ in \figpanel{fig_model}{g}, we see that the quantum Zeno crossover persists almost unchanged with this drive. Compared to the case $\tilde{\Omega}=0$, the crossover point is slightly increased and $\Gamma'_{2, \text{eff}}$ is reduced.

Another possible twist to this model is to consider post-selection. Recently, the technique of selecting particular quantum paths in a time evolution by discarding others via post-selection~\cite{Li2018, Naghiloo2019, Skinner2019} has attracted much interest. In particular, the dynamics of a system on selected paths with no quantum jumps can be described by an effective non-Hermitian Hamiltonian. For the Liouvillian of the two-qubit system here in \eqref{eq_model_L}, the effective Hamiltonian in the frame rotating at the resonant qubit frequencies is (see \appref{app_H})
\beq \label{eq_model_Heff}
\tilde{H}_\text{eff} = \tilde{g} \mleft( \sigma_1^+ \sigma_2^- + \text{H.c.} \mright) + \tilde{\Omega} \sigma_1^x - i \frac{\tilde{\Gamma}}{4} \mleft(\sigma^z_1 + \mathbf{I}\mright),
\eeq
where $\mathbf{I}$ is the identity matrix. For $\tilde{\Omega}=0$, the evolution of $n_2(\tilde{t})$ shows a transition from oscillatory to non-oscillatory dynamics at $\tilde{\Gamma} = 3.8 \tilde{g}$; see \figpanel{fig_model}{h}. Just like for the quantum Zeno crossover above, the deviation from the transition point $\tilde{\Gamma} = 4 \tilde{g}$
~\cite{Minganti2019} is due to the finite time considered. We note that this kind of transition has been observed in experiment in a similar model on a single qubit~\cite{Naghiloo2019}.

For a quantum simulator to simulate the above models and characterize the quantum Zeno crossover in the Liouvillian dynamics and the transition from oscillatory to non-oscillatory dynamics in the effective Hamiltonian dynamics, it needs to be versatile when it comes to tuning the ratio $\Gamma'/g'$. In the next section, we show in detail how the giant-atom quantum simulator illustrated in \figref{fig_simulator} in \secref{sec_sim_idea} achieves this tunability.


\section{Quantum Zeno crossover in a giant-atom quantum simulator}
\label{Ch_sim}

In \secref{sec_sim_idea}, we described how a giant-atom quantum simulator consisting of two giant atoms can reach a wide range of different parameter regimes solely by tuning the transition frequencies of the atoms. In this section, we show the details of how to harness this tunability in practice to efficiently simulate the Liouvillian and effective non-Hermitian Hamiltonian dynamics for the two-qubit model system introduced in \secref{Ch_model}. We provide a concrete simulation protocol and characterize its performance.


\subsection{Liouvillian dynamics}
\label{sec3.1}

\begin{figure*}
\center
\includegraphics[width=\linewidth]{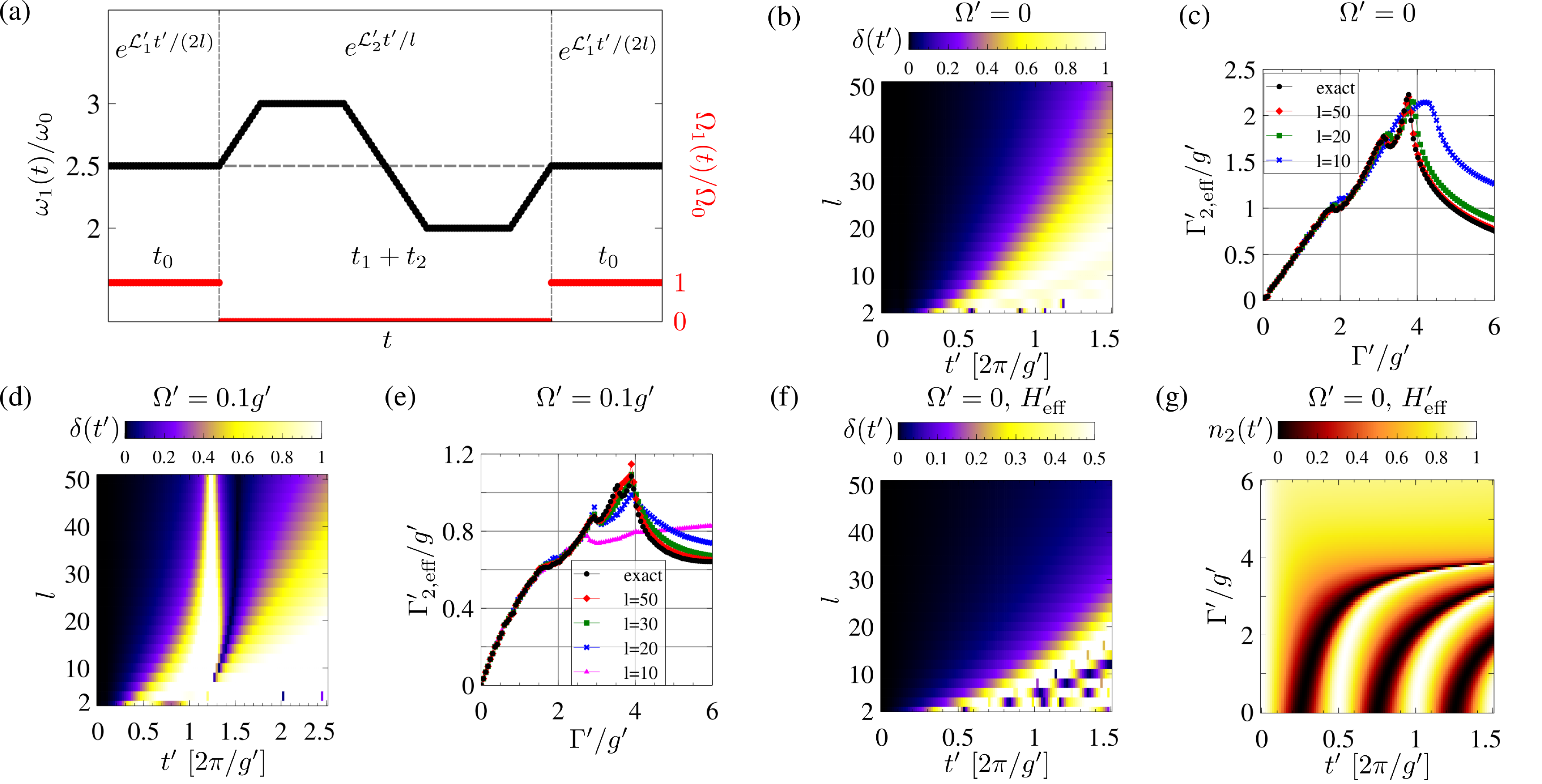}
\caption{The giant-atom quantum simulation protocol and its numerically simulated results for the two-qubit model from \secref{Ch_model}.
(a) The protocol for tuning giant-atom parameters during one Trotter step in the simulation of \eqref{eq_model_L} using the giant-atom quantum simulator shown in \figref{fig_simulator}. During the Trotter step, $\omega_1$ and $\Omega_1$ are tuned as shown ($\omega_0 = 2 \pi v / (\Delta x_1 + \Delta x_2)$ as discussed below \eqref{eq_simulator_L} and $\Omega_0 = \tilde{\Omega} g_0 / \tilde{g}$) while $\omega_2 = \omega_\text{DF}$ and $\Omega_2 = 0$ remain fixed.
(b) The simulation error $\delta(\tilde{t})$ [defined in \eqref{eq_err}] for the Liouvillian dynamics with $\Omega' = 0$ and $\tilde{\Gamma} = 6 \tilde{g}$.
(c) The simulated effective decay rate $\Gamma'_{2, \text{eff}}$ for the Liouvillian dynamics with $\Omega' = 0$. The simulation with $l = 50$ Trotter steps predicts the quantum Zeno crossover point very well.
(d) The simulation error for the Liouvillian dynamics with $\Omega' = 0.1 g'$ and $\tilde{\Gamma} = 6\tilde{g}$.
(e) The simulated effective decay rate $\Gamma'_{2, \text{eff}}$ for the Liouvillian dynamics with $\Omega' = 0.1g'$. The simulation with $l = 50$ Trotter steps predicts the quantum Zeno crossover point very well.
(f) The simulation error for the effective non-Hermitian Hamiltonian dynamics with $\Omega' = 0$ and $\tilde{\Gamma} = 6 \tilde{g}$. The error is smaller than that for the simulation of the corresponding Liouvillian dynamics in panel (b).
(g) The simulated effective non-Hermitian Hamiltonian dynamics with $l = 30$ Trotter steps, which faithfully reproduces the transition from oscillatory to non-oscillatory dynamics.}
\label{fig_sim}
\end{figure*}

Since the Liouvillian of the two-qubit system we wish to simulate [\eqref{eq_model_L}] can be split into two terms 
\begin{align}
\tilde{\mathcal{L}}_1 \rho &= -i \mleft[ \tilde{g} \mleft( \sigma_1^+ \sigma_2^- + \text{H.c.} \mright) + \tilde{\Omega} \sigma_1^x , \rho \mright] , \\
\tilde{\mathcal{L}}_2 \rho &= -\tilde{\Gamma} \mathcal{D}[\sigma_1^-] \rho , 
\end{align}
we can decompose its dynamics using the second-order Trotter--Suzuki decomposition~\cite{SUZUKI1990319, Kliesch2011, Werner2016, Childs2017}:
\begin{align}
\exp \mleft( \tilde{\mathcal{L}} \tilde{t} \mright) &= \mleft[ \exp \mleft( \frac {\tilde{\mathcal{L}}_1 \tilde{t}}{2l} \mright) \exp \mleft( \frac{\tilde{\mathcal{L}}_2 \tilde{t}}{l} \mright) \exp \mleft( \frac{\tilde{\mathcal{L}}_1 \tilde{t}}{2l} \mright) \mright]^l \nonumber \\
&\quad + O\mleft(\frac{\tilde{t}^3}{l^2}\mright).
\label{eq_Trotter}
\end{align}
The coherent dynamics generated by $\exp \mleft[ \tilde{\mathcal{L}}_1 \tilde{t} / (2l) \mright]$ can be simulated by setting $\omega_1 = \omega_2 = \omega_\text{DF}$ [to have qubit-qubit coupling $g_0$ while $\Gamma_k = 0$ and $\Gamma_\text{coll} = 0$; see \figpanel{fig_simulator}{b}] and $\Omega_1 = \tilde{\Omega} g_0 / \tilde{g}$, and letting the system evolve for a time $t_0 = \tilde{g} \tilde{t} / (2 g_0 l)$. The dynamics generated by $\exp \mleft( \tilde{\mathcal{L}}_2 \tilde{t} / l \mright)$ is simply the decay of qubit 1 at a rate $\tilde{\Gamma}$ for a time $\tilde{t} / l$. This decay can be simulated by fixing $\omega_2 = \omega_\text{DF}$ and tuning $\omega_1$ to a frequency where qubit 1 decays; see \figpanel{fig_simulator}{b}.

We thus need to tune the frequency of qubit 1 back and forth between different values. When doing so, it is crucial to align the phase between the two qubits such that the next Trotter step provides correct dynamics. We therefore tune $\omega_1$ symmetrically around $\omega_\text{DF}$:
\begin{widetext}
\beq \label{eq_omega}
\omega_1(t) = \mleft\{
\begin{array}{crl}
\omega_\text{DF} + v_1(t-t_0) & \qquad t - t_0 &\hspace{-0.1cm}< \frac{t_1}{4} \\[0.1cm] 
\omega_\text{DF} + v_1 \frac{t_1}{4} & \qquad \frac{t_1}{4} < t - t_0 &\hspace{-0.1cm}< \frac{t_1}{4} + \frac{t_2}{2} \\[0.1cm] 
\omega_\text{DF} - v_1 \mleft( t - t_0 - \frac{t_1 + t_2}{2} \mright) & \quad \frac{t_1}{4} + \frac{t_2}{2} < t - t_0 &\hspace{-0.1cm}< \frac{3 t_1}{4} + \frac{t_2}{2} \\[0.1cm] 
\omega_\text{DF} - v_1 \frac{t_1}{4} & \quad \frac{3 t_1}{4} + \frac{t_2}{2} < t - t_0 &\hspace{-0.1cm}< \frac{3 t_1}{4} + t_2 \\[0.1cm] 
\omega_\text{DF} + v_1 \mleft( t - t_0 - t_1 - t_2 \mright) & \quad \frac{3 t_1}{4} + t_2 < t - t_0 &\hspace{-0.1cm}< t_1 + t_2 \\[0.1cm] 
\omega_\text{DF} & \text{otherwise} ,
\end{array}
\mright.
\eeq
\end{widetext}
where $v_1$ is the speed of the frequency change of the qubit and the times $t_{1,2}$ are determined by $\int_0^{t_1 + t_2} \Gamma_1[\omega_1(t)] dt = \tilde{\Gamma} \tilde{t} / l$ (see \appref{app_omega} for the full derivation). The time dependence of $\omega_1$ and $\Omega_1$ during one Trotter step are shown in \figpanel{fig_sim}{a}; $\omega_2 = \omega_\text{DF}$ and $\Omega_2 = 0$ remain fixed throughout the whole simulation. The total simulated time-evolution operator after $l$ Trotter steps is given by
\beq
\mleft[ \exp \mleft( \tilde{\mathcal{L}} \tilde{t} \mright) \mright]_\text{sim} = \mleft( \exp \mleft[ \int_0^{2 t_0 + t_1 + t_2} \mathcal{L}(t) dt \mright]\mright)^l ,
\eeq
such that $\rho_\text{sim}(\tilde{t}) = \mleft[ \exp \mleft( \tilde{\mathcal{L}} \tilde{t} \mright) \mright]_\text{sim} \rho(0)$.

We are now ready to numerically simulate our quantum-simulation scheme. For concreteness, we consider parameters that are experimentally accessible for superconducting qubits: $\Delta x_1 + \Delta x_2 = \qty{8.125}{\cm}$~\cite{Kannan2020, Sundaresan2015} and $v = \qty{1.3e8}{\meter/\second}$~\cite{Goppl2008, Blais2021}; these together yield $\omega_0 / (2\pi) = \qty{1.6}{\giga\hertz}$ and thus $\omega_\text{DF} / (2\pi) = \qty{4.0}{\giga\hertz}$. This relatively small value of $\omega_0$ helps prevent excessive coupling to other environments such as the readout resonators by giving a large detuning of the qubits~\cite{Blais2021}. We set the speed of changing qubit 1's frequency to $v_1 / (2\pi) = \qty{0.2}{\giga\hertz/\nano\second}$, such that the time $t_1$ spent to tune the qubit frequency is not so large. Finally, we set the qubit-waveguide coupling to $\gamma / (2\pi) = \qty{1}{\mega\hertz}$. We note that the value of $\gamma$ does not influence the simulation result if the qubits do not couple to other environments beyond the waveguide (as we assume in this Section), since it is only the ratio $\gamma / \Omega_1$ that needs to be tuned and we easily can choose $\Omega_1$ in a wide range spanning several orders of magnitude. In realistic cases, some coupling to other environments is inevitable; we analyze the effects of such imperfections in \secref{Ch_err}.

In order to characterize the quantum Zeno crossover, a faithful simulation of the population $n_2(\tilde{t})$ of qubit 2 is essential. In particular, since the effective decay rate $\Gamma'_{2, \text{eff}}$ of that qubit is determined by $n_2(\tilde{t})-n_2(\tilde{t}\to\infty)$, the error in $n_2(\tilde{t})$ should not be too large compared to this value. We therefore define the simulation error as~\footnote{We note that the usual definition of fidelity of quantum states, $\mathcal{F}(\rho_\text{sim}(t), \rho(t)) = \mleft[ \text{Tr} \sqrt{\sqrt{\rho_\text{sim}(t)} \rho(t) \sqrt{\rho_\text{sim}(t)}} \mright]^2$, is not a good measure of the simulation accuracy for our purposes, since it is not sensitive to the errors in $n_2(t)$. In particular, when $n_2(t)$ is small, $\mathcal{F}(\rho_\text{sim}(t), \rho(t))$ can remain close to 1 even when the normalized population deviation $\delta(t)$ in \eqref{eq_err} is large.}
\beq \label{eq_err}
\delta(\tilde{t}) = \frac{\mleft| n_{2,\text{sim}}(\tilde{t}) - n_2(\tilde{t}) \mright|}{n_2(\tilde{t}) - n_2(\tilde{t} \to \infty)} ,
\eeq
where $n_{2,\text{sim}}(\tilde{t}) = \mleft\{ 1 + \text{Tr} \mleft[ \sigma_2^z \rho_\text{sim}(\tilde{t}) \mright] \mright\} / 2$ is the population of qubit 2 obtained in the simulation, which can be directly measured in an actual experiment. For $\tilde{\Omega}=0$, we have $n_2(\tilde{t}\to\infty)=0$; for $\tilde{\Omega} = 0.1 \tilde{g}$, $n_2(\tilde{t} \to \infty)$ is computed in \appref{app_L}.

We begin with the case of no drive, i.e., $\tilde{\Omega} = 0$. In \figpanel{fig_sim}{b}, we plot the simulation error $\delta$ as a function of $\tilde{t}$ and $l$ for $\tilde{\Gamma} = 6 \tilde{g}$. The result is similar for other values of $\tilde{\Gamma}$. We see that to maintain a constant simulation error, $l$ must scale superlinearly with $\tilde{t}$, which is in agreement with the scaling of the Trotter error in \eqref{eq_Trotter}. Next, we show, in \figpanel{fig_sim}{c}, the fitted effective decay rate $\Gamma'_{2, \text{eff}}$ from simulation results obtained with different numbers $l$ of Trotter steps. We observe that, for small $l$, a significant error in $\Gamma'_{2, \text{eff}}$ mainly appears when $\tilde{\Gamma}$ is large. We also note that, for $l = 20$, the simulated dynamics predicts the quantum Zeno crossover point at the same value as the exact dynamics. The main advantage of going to larger $l$ is thus that the effective decay rates can be predicted more accurately.

In the case $\tilde{\Omega}=0.1\tilde{g}$, we observe similar behavior for the Trotter error as without drive; see \figpanel{fig_sim}{d}. The main difference compared to \figpanel{fig_sim}{b} is an increase of $\delta(\tilde{t})$ around $\tilde{t} = 2.6 \pi / \tilde{g}$. This increase is due to oscillations in $n_2(\tilde{t})$: $n_2(\tilde{t}) - n_2(\tilde{t} \to \infty)$ approaches 0 around $\tilde{t} = 2.4 \pi / \tilde{g}$ and then increases again. The fitted effective decay rate from the simulated dynamics faithfully captures the reduction compared to the case of $\tilde{\Omega}=0$; see \figpanel{fig_sim}{e}. With $l = 30$ Trotter steps, the quantum Zeno crossover is predicted well. The reason for needing a larger $l$ than in the case of $\tilde{\Omega} = 0$ is that the maximum simulation time is larger here. 

The results displayed in \figpanels{fig_sim}{b}{e} demonstrate the capability of the giant-atom quantum simulator to simulate the dynamics of the two-qubit model from \secref{Ch_model}. In particular, using realistic experimental parameters, we see that relatively few Trotter steps sufficed to correctly characterize the quantum Zeno crossover in this model.


\subsection{Effective non-Hermitian Hamiltonian dynamics}

As discussed at the end of \secref{Ch_model}, post-selecting the instances of Liouvillian dynamics without quantum jumps yields dynamics that can be described by an effective non-Hermitian Hamiltonian. In our giant-atom quantum simulator, such post-selection can be performed in at least two ways: by detecting photons emitted into the waveguide~\cite{Gu2017, Kono2018, Besse2018, Lescanne2020} or by measuring the total population in the giant atoms~\cite{Naghiloo2019, Hoke2023}. For the microwave photons in superconducting circuits, the latter method appears generally easier and more precise. In particular, it has been demonstrated with at least 12 qubits~\cite{Hoke2023}. If no photons are detected in the waveguide during the whole dynamics, or the total qubit population is unchanged (for cases without any drive), we can conclude that no quantum jump has occurred. 

For the example with two giant atoms considered here, the dynamics under post-selection of the giant-atom quantum simulator are given by 
\beq
\rho(t) = \exp \mleft( - i H_\text{eff} t \mright) \rho(0) \exp \mleft( i H^\dag_\text{eff}t \mright)
\eeq
with (see \appref{app_H} for the full derivation)
\begin{align}
H_\text{eff}(t) &= g(\omega_1,\omega_2) \mleft( \sigma_1^+ \sigma_2^- + \text{H.c.} \mright) + \omega_1 \frac{\sigma_1^z}{2} + \omega_2  \frac{\sigma_2^z}{2}\nonumber\\
&\quad + \Omega_1(t) \sigma_1^x + \Omega_2(t) \sigma_2^x\nonumber\\
&\quad -i \frac{\Gamma_1(\omega_1)}{4} \mleft( \sigma_1^z + \mathbf{I} \mright) - i \frac{\Gamma_2(\omega_2)}{4} \mleft( \sigma_2^z + \mathbf{I} \mright) \nonumber\\
&\quad -i \frac{\Gamma_\text{coll}(\omega_1,\omega_2)}{2} \mleft( \sigma_1^+ \sigma_2^- + \text{H.c.} \mright).
\label{eq_sim_Heff}
\end{align}
Using the same protocol for tuning the giant-atom parameters [see \figpanel{fig_sim}{a}] as for the Liouvillian case in \secref{sec3.1}, we can simulate the effective Hamiltonian in \eqref{eq_model_Heff}. 

The results of this simulation are shown in \figpanel{fig_sim}{f, g}. Similar to the Liouvillian case, we see in \figpanel{fig_sim}{f} that to keep the Trotter error constant, the number of Trotter steps $l$ has to scale super-linearly with $t'$. For $l=30$, the simulated dynamics shown in \figpanel{fig_sim}{g} predicts the transition to be at $\tilde{\Gamma}\approx 3.9 \tilde{g}$, just \qty{2.5}{\percent} from the exact result in \figpanel{fig_model}{h}. 

These results demonstrate the capability of the giant-atom quantum simulator to simulate the effective non-Hermitian Hamiltonian dynamics of the two-qubit model. In particular, an experimentally feasible small number of Trotter steps is sufficient to characterize the transition from oscillatory to non-oscillatory dynamics in this model.


\section{Potential simulation errors from noise and other imperfections}
\label{Ch_err}

In the preceding section, we saw how the Trotterization of the dynamics introduces some errors in the quantum simulation. Those errors can be reduced by decreasing the length of the Trotter steps (thus increasing their number $l$). In this section, we discuss and analyze other potential error sources for our quantum simulation scheme. 

The impact of various errors on a quantum simulation will in many cases depend on both the system that one aims to simulate and the protocol used to carry out the simulation~\cite{Hauke2012}. This situation is similar to how knowing individual gate errors in a quantum computer does not mean that one knows how an algorithm will perform when implemented using those gates~\cite{Lagemann2023}. In general, the aim in the era of noisy intermediate-scale quantum (NISQ) devices is to find problems where quantum simulators can determine some quantity that is robust to errors, yet hard for a classical simulator to calculate~\cite{Preskill2018quantumcomputingin}.


In the setups with superconducting giant artificial atoms that we consider, the main cause of realistic imperfections is the coupling of the qubits to other environments than the waveguide, e.g., the readout resonator and its surroundings or two-level systems within the qubit material~\cite{Krantz2019, Muller2019}. Such couplings can result in additional decay and dephasing of the qubits, which is a typical technical challenge in the NISQ era~\cite{Preskill2018quantumcomputingin, Krantz2019}. The simulation errors caused by this additional noise in the qubits are protocol- and problem-dependent. Below, we show how these errors influence the performance of the giant-atom quantum simulator for the two-qubit model from \secref{Ch_model}. Since the only energy scale that enters the dynamics for those simulations is the qubit-waveguide coupling $\gamma$, the threshold where extra decay at a rate $\Gamma_\text{ex}$ and extra dephasing at a rate $\Gamma_\phi$ adversely impacts the prediction of the Zeno or oscillatory-to-non-oscillatory crossover are given in units of that coupling.

We note that furthermore, statistical errors resulting from an insufficient amount of repeated experiments, or imperfect post-selection due to insufficiently sensitive photon detectors, can also increase the error in the giant-atom quantum simulator. A quantitative analysis of the influence of these imperfections is given in \appref{sec_app_err}. We do not analyze the impact of other relatively small potential errors, such as the potential distortion of the qubit control signals due to insufficient characterization of the transfer function for the qubit control lines~\cite{Jerger2019}.


\subsection{Effect of extra decay}
\label{sec:extra_decay}

We first consider how extra decay to some environment other than the waveguide affects the quantum-simulation results from \secref{Ch_sim}. We assume that this extra decay occurs at a rate $\Gamma_\text{ex}$ for both qubit 1 and qubit 2, such that a term $\Gamma_\text{ex} \mleft( \mathcal{D}[\sigma_1^-] + \mathcal{D}[\sigma_2^-] \mright)\rho$ is added to the right-hand side of \eqref{eq_simulator_L}.

\begin{figure}
\center
\includegraphics[width=\linewidth]{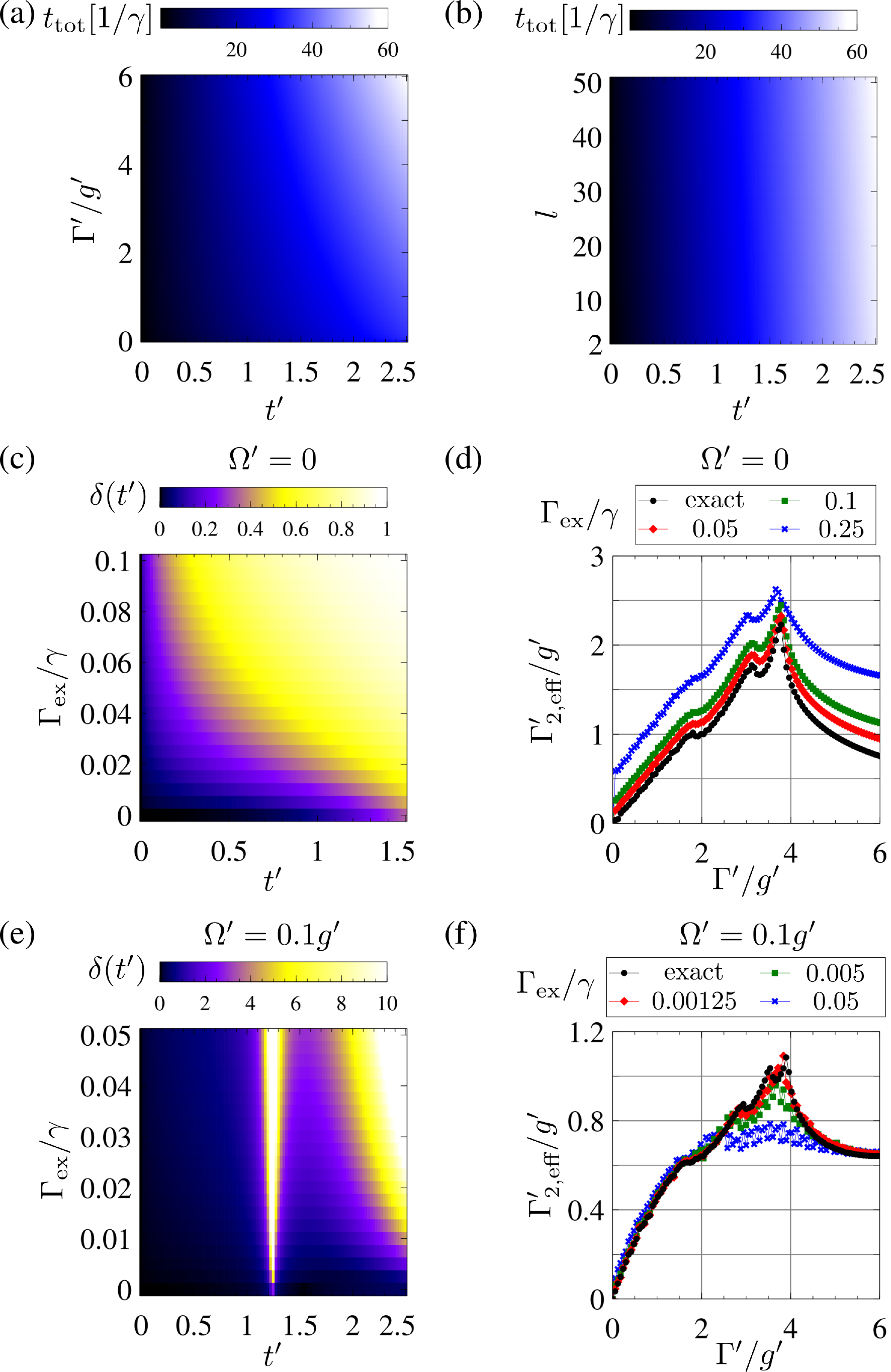}
\caption{Errors induced in the quantum simulation of Liouvillian dynamics by finite qubit lifetimes $1 / \Gamma_\text{ex}$.
(a,b) The total simulation time $t_\text{tot}$ as a function of $\tilde{\Gamma}$, $\tilde{t}$, and $l$. In panel (a), we fix $l=50$; in panel (b), we fix $\tilde{\Gamma} = 6 \tilde{g}$.
(c) Simulation error $\delta(t')$ [see \eqref{eq_err}] as a function of $t'$ and $\Gamma_\text{ex}$ with $\Omega' = 0$, $l = 50$, and $\Gamma' = 4 g'$.
(d) Simulated effective relaxation rate $\Gamma'_{2, \text{eff}}$ as a function of $\Gamma'$ for various values of $\Gamma_\text{ex}$, with $\Omega' = 0$.
(e) $\delta(t')$ as a function of $t'$ and $\Gamma_\text{ex}$ with $\Omega' = 0.1 g'$, $l = 50$, and $\Gamma' = 6 g'$.
(f) $\Gamma'_{2, \text{eff}}$ as a function of $\Gamma'$ for various values of $\Gamma_\text{ex}$, with $\Omega' = 0.1 g'$.}
\label{fig_err}
\end{figure}

Since the simulation error in our case depends on both the decay rate $\Gamma_\text{ex}$ and the total simulation time $t_\text{tot} = l (2 t_0 + t_1 + t_2)$, we first remind ourselves how $t_\text{tot}$ is connected to the simulated relaxation rate $\tilde{\Gamma}$, the simulated time $\tilde{t}$, and the number of Trotter steps $l$. As shown in \figpanel{fig_err}{a}, $t_\text{tot}$ increases as $\tilde{\Gamma}$ and $\tilde{t}$ increases. This behaviour is expected since $t_1 + t_2$ increases with $\tilde{\Gamma}$, and both $t_0$ and $t_1 + t_2$ increase with $\tilde{t}$. However, as shown in \figpanel{fig_err}{b}, $t_\text{tot}$ is not significantly affected by $l$. We therefore fix $l = 50$ when analyzing the impact of extra decay.

In \figpanel{fig_err}{c}, we show the simulation error $\delta(\tilde{t})$ as a function of $\Gamma_\text{ex}$ and $t'$ for $\Omega' = 0$ and $\Gamma' = 4 g'$. We see that the simulation error increases with both $\Gamma_\text{ex}$ and $t'$. The results are similar for other choices of $\Gamma'$, e.g., for $\Gamma_\text{ex} = 0$, we have almost the same results as at the top of \figpanel{fig_sim}{b}, where $\Gamma' = 6 g'$.  

We next look at the effect on the simulated effective relaxation rate $\Gamma'_{2, \text{eff}}$ in \figpanel{fig_err}{d}. While $\Gamma'_{2, \text{eff}}$ increases with $\Gamma_\text{ex}$ and thus increasingly deviates from the correct value, this does not significantly influence the location of the quantum Zeno crossover point in the simulation. This crossover point appears quite robust to extra decay in the simulator qubits up to at least $\Gamma_\text{ex} = 0.1 \gamma$. For a conservatively low choice of qubit-waveguide coupling of $\gamma/(2\pi) = \qty{1}{\mega\hertz}$, that level of extra decay corresponds to a qubit lifetime of $\qty{1.6}{\micro\second}$, which is much smaller than the current state-of-the-art of several hundred microseconds~\cite{annurev_qubits, Place2021, Somoroff2023, Kim2023, Biznarova2023, Kono2024, Bal2024}. 

We also consider a case with nonzero simulated driving: $\tilde{\Omega} = 0.1 g'$. Setting $\Gamma' = 6 g'$ again, we show the simulation error for this case in \figpanel{fig_err}{e}. Here, a large error appears in the simulation around $t' = 2.4 \pi / g'$. The reason for this error is the same as in \figpanel{fig_sim}{f}: $\tilde{\Omega} \neq 0$ results in oscillations in $n_2(t')$, and near this particular $t'$, $n_2(t')\approx n_2(t'\to\infty)$, such that the denominator in \eqref{eq_err} approaches zero.

Compared to $\tilde{\Omega} = 0$, the error due to extra decay for $\tilde{\Omega} = 0.1 g'$ is significantly increased. As shown in \figpanel{fig_err}{f}, the predicted effective relaxation rate $\Gamma'_{2, \text{eff}}$ has a large relative error when $\tilde{\Gamma}$ is close to the quantum Zeno crossover point, already for $\Gamma_\text{ex} = 0.05 \gamma$. Going down to $\Gamma_\text{ex} = 0.005\gamma$, the error in the effective relaxation rate becomes small, but the prediction of the quantum Zeno crossover point is clearly larger than it was for $\tilde{\Omega} = 0$ in \figpanel{fig_err}{d}. To obtain a good agreement with the quantum Zeno crossover point predicted by the exact evolution, the rate of extra decay cannot be larger than around $\Gamma_\text{ex}= 1.25 \cdot 10^{-3} \gamma$. For $\gamma/(2\pi) = \qty{1}{\mega\hertz}$, this extra decay translates into a qubit lifetime larger than $\qty{127}{\micro\second}$, which still is within the limit of state-of-the-art experiments. Furthermore, this requirement on the extra decay can be softened by considering a larger $\gamma$, as long as $\gamma \ll \omega_{1,2}$ such that the Markovian approximation is valid.

Finally, let us comment on the simulation of effective non-Hermitian Hamiltonian dynamics. Unlike the simulation of Liouvillian dynamics, such a simulation is not influenced by extra decay for the parameters we considered here. The reason for this robustness is that in the effective non-Hermitian Hamiltonian dynamics, the total qubit population $n_1 + n_2$ is conserved due to the absence of quantum jumps (and drive). For the case $n_1 + n_2 = 1$ that we consider here, the extra decay term that gets added to \eqref{eq_sim_Heff} is proportional to identity, and thus does not influence the dynamics. The only effect of the extra decay will be that more experiments are required before enough trajectories without quantum jumps are registered. If the extra decay rates for the two qubits differ, the cancellation in \eqref{eq_sim_Heff} will not be perfect, and there will be some error in the quantum simulation due to the extra decay.


\subsection{Effect of extra dephasing}

We now turn to the effect of extra dephasing on the quantum simulation. We assume that this extra dephasing occurs at the same rate $\Gamma_\phi$ for both qubits, such that it is captured by adding the term $\mleft( \Gamma_\phi / 2 \mright) \mleft( \mathcal{D}[\sigma_1^z] + \mathcal{D}[\sigma_1^z] \mright)\rho$ to the right-hand side of \eqref{eq_simulator_L}.

\begin{figure*}
\center
\includegraphics[width=\linewidth]{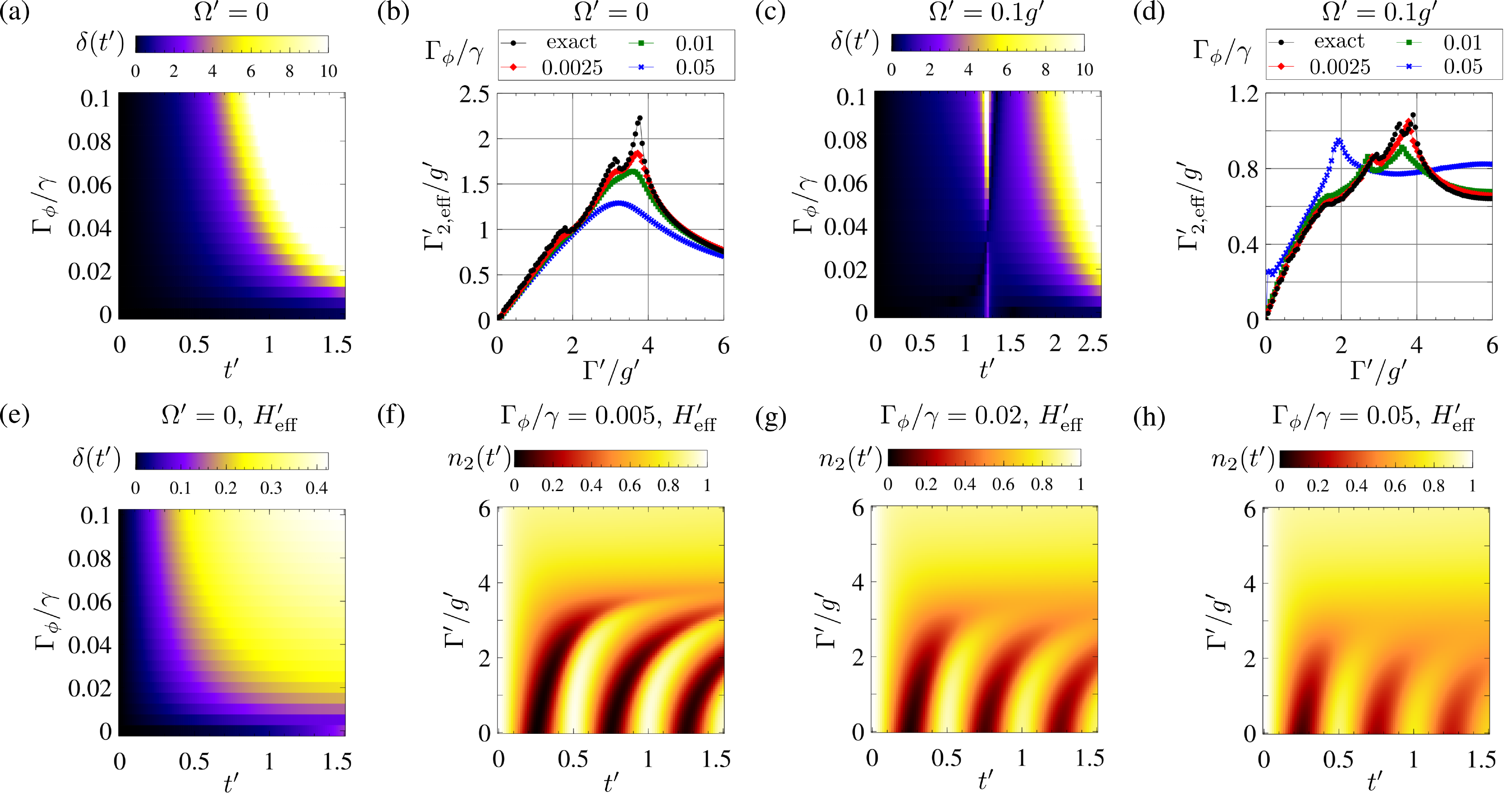}
\caption{Simulation error induced by a finite qubit dephasing time $1 / \Gamma_\phi$ in the quantum simulation of (a-d) Liouvillian dynamics and (e-h) effective non-Hermitian Hamiltonian dynamics.
(a) Simulation error $\delta(t')$ [see \eqref{eq_err}] as a function of qubit dephasing rate $\Gamma_\phi$ for $\Omega' = 0$ with $\Gamma' = 4 g'$ and $l = 50$.
(b) Simulated effective relaxation rate $\Gamma'_{2, \text{eff}}$ as a function of $\Gamma_\phi$ for $\Omega'=0$.
(c) $\delta(t')$ as a function of $\Gamma_\phi$ for $\Omega' = 0.1g'$ with $\Gamma' = 4 g'$ and $l = 50$.
(d) $\Gamma'_{2, \text{eff}}$ as a function of $\Gamma_\phi$ for $\Omega' = 0.1 g'$.
(e) $\delta(t')$ as a fucntion of $\Gamma_\phi$ for the effective non-Hermitian Hamiltonian dynamics with $\Omega' = 0$, $\Gamma' = 4 g'$, and $l = 30$.
(f)-(h) Simulations of the effective non-Hermitian Hamiltonian dynamics for three different $\Gamma_\phi$.}
\label{fig_err_dephasing}
\end{figure*}

We first consider the effects of dephasing on the quantum simulation of the Liouvillian dynamics. As shown in \figpanel{fig_err_dephasing}{a} for $\Omega' = 0$, we find that dephasing causes a much larger simulation error than extra decay does [compare \figpanel{fig_err}{c}]. We attribute this relative increase in simulation error for dephasing to the fact that the quantum simulation requires phase alignment of the qubits to perform two-qubit XY-gates in the Trotter steps, and the dephasing impacts this phase alignment.

Furthermore, as shown in \figpanel{fig_err_dephasing}{b}, significant errors in the simulated effective relaxation rate $\Gamma'_{2, \text{eff}}$ appear at lower extra dephasing rates than extra decay rates, and the location of the quantum Zeno crossover point is not as robust to extra dephasing as it is to extra decay [compare \figpanel{fig_err}{d}]. Indeed, we see in \figpanel{fig_err_dephasing}{b} that the dephasing should not exceed about $2.5 \cdot 10^{-3} \gamma$ if the crossover point is to be simulated correctly. For $\gamma/(2\pi)=\qty{1}{\mega\hertz}$, this threshold value for the dephasing is $\Gamma_\phi/(2\pi) \approx \qty{2.5}{\kilo\hertz}$.

Adding a drive term with $\Omega' = 0.1g'$, we see in \figpanel{fig_err_dephasing}{c} that the simulation error is not increased compared to $\Omega' = 0$. Also, the requirement for obtaining a faithful quantum Zeno crossover point is similar to that for $\Omega'=0$ [\figpanel{fig_err_dephasing}{d}]. The dephasing threshold for obtaining a faithful simulation result for the crossover point increases linearly with $\gamma$; with $\gamma / (2\pi) = \qty{10}{\mega\hertz}$, the threshold becomes $\Gamma_\phi / (2 \pi) \approx \qty{25}{\kilo\hertz}$, which can be achieved in state-of-the-art tunable qubits~\cite{annurev_qubits, lacroix2023fast}.

For the effective non-Hermitian Hamiltonian dynamics, there is no mitigating cancellation effect of errors as there was for extra decay (see \secref{sec:extra_decay}). Instead, the dephasing yields simulation errors [see \figpanel{fig_err_dephasing}{e}] due to the breakdown of phase alignment of the qubits. In particular, the stronger the dephasing, the smaller the oscillation amplitude in the simulated dynamics, which hinders the transition from oscillatory to non-oscillatory dynamics, as shown in \figpanels{fig_err_dephasing}{f}{h}. Qualitatively, the transition from oscillatory to non-oscillatory dynamics remains visible in the right place for $\Gamma_\phi$ up to around $0.005 \gamma$.


\section{Scaling up the giant-atom quantum simulator for driven-dissipative spin chains} \label{Ch_scale_up}

Having seen in detail how the giant-atom quantum simulator works for a two-qubit example, we now turn to discuss how such a simulator can be scaled up to simulate large open quantum many-body systems. We begin by showing how giant atoms can simulate a one-dimensional driven-dissipative spin chain with nearest-neighbor interactions. We then show that by rearranging the coupling points of the giant atoms, we can extend this setup to simulate driven-dissipative spin chains with long-range (even all-to-all) interactions. We end this section with a discussion of potential limitations to scaling up a giant-atom quantum simulator.


\subsection{Simulation of driven-dissipative spin chains with nearest-neighbor interactions}
\label{sec:nearest-neighbor}

\begin{figure*}
\center
\includegraphics[width=\linewidth]{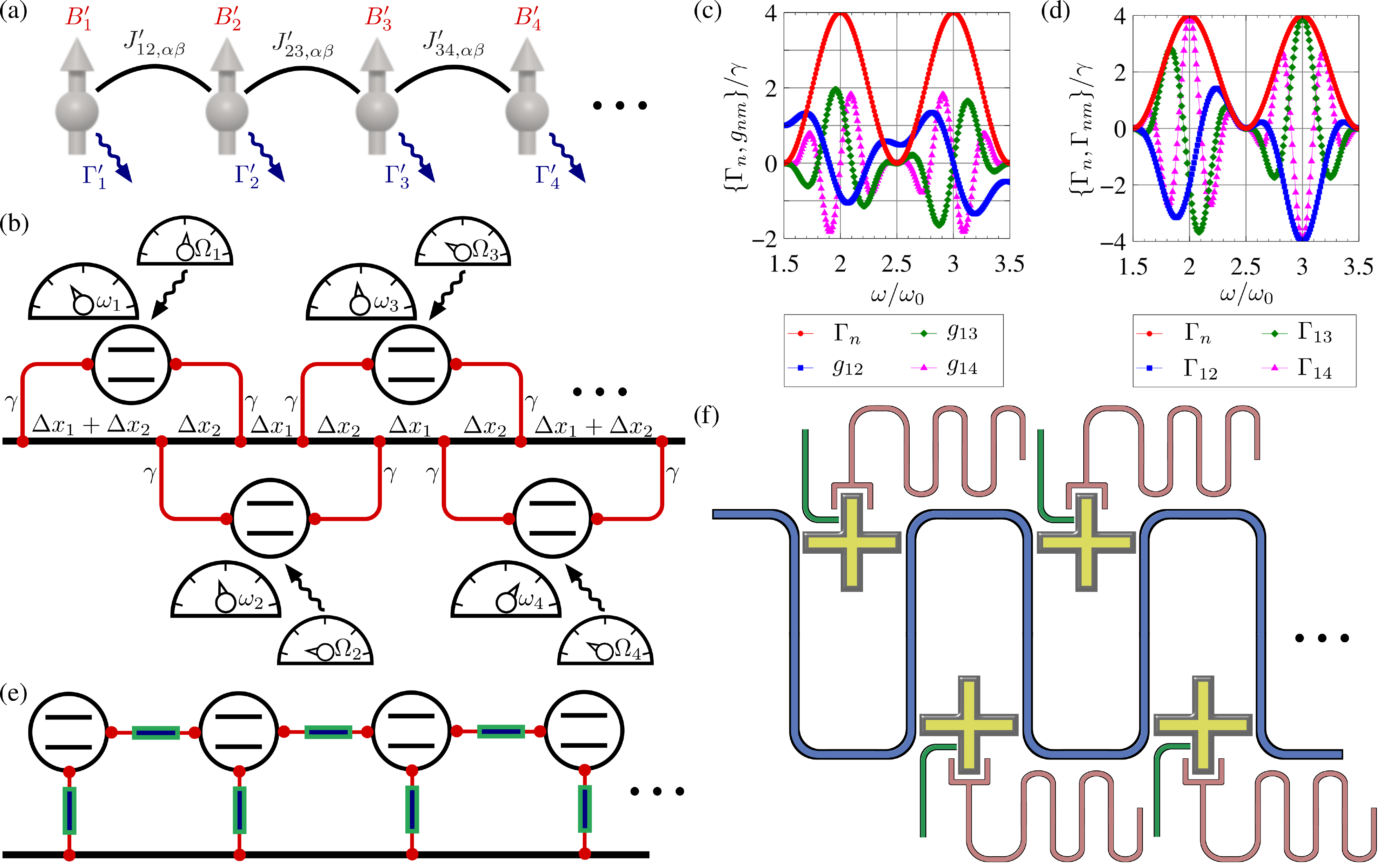}
\caption{A giant-atom quantum simulator for the simulation of driven-dissipative spin chains with nearest-neighbor interactions.
(a) A sketch of the driven-dissipative spin chain model [\eqref{eq:L_model_NN_spin}] to be simulated.
(b) A setup for the giant-atom quantum simulator, with neighboring giant atoms coupling to the waveguide in a braided configuration.
(c,d) Qubit-qubit couplings $g_{nm}$ and decay rates $\Gamma_n$ and $\Gamma_{nm}$ as functions of qubit frequencies $\omega$ for the simulator in panel (b) with four giant atoms and $\Delta x_1 = 4 \Delta x_2$.
(e) A small-atom quantum simulator to simulate the model in panel (a), where parametric couplers (blue rectangles with green edges) are used to tune the parameters in panels (c) and (d).
(f) Sketch of a possible experimental setup for the giant-atom quantum simulator with superconducting qubits (yellow) coupled to the waveguide (blue). Flux lines (green) are used to tune the qubit frequencies and resonators (pink) are coupled to the qubits to enable single-qubit drives and measurements.}
\label{fig_4_atom_1}
\end{figure*}

We first consider the quantum simulation of a driven-dissipative spin chain with only nearest-neighbor interactions, as illustrated in \figpanel{fig_4_atom_1}{a}. The Liouvillian and effective non-Hermitian Hamiltonian dynamics for such a system are given by
\begin{align} 
&\tilde{H} = \sum_{n, \alpha, \beta} \tilde{J}_{n, n+1, \alpha \beta} S_n^\alpha S_{n+1}^\beta + \sum_n \tilde{B}_n S^x_n , \label{eq:H_model_NN_spin} \\
&\tilde{\mathcal{L}} \rho = - i \mleft[ \tilde{H}, \rho \mright] + \sum_{n} \tilde{\Gamma}_n \mathcal{D}[\sigma_n^-]\rho , \label{eq:L_model_NN_spin} \\
&\tilde{H}_\text{eff} = \tilde{H} - i \sum_n \frac{\Gamma'_n}{4} \sigma^z_n, \label{eq:Heff_model_NN_spin}
\end{align}
where $n$ is the site index, and $\alpha, \beta$ are spin components. Models described by \eqref{eq:L_model_NN_spin} include Ising models~\cite{Cai2013, gustafson2023quantum} and XXZ models~\cite{Cai2013, Yamamoto2022, Chen2023} subject to onsite dissipation.

The dynamics of Eqs.~(\ref{eq:L_model_NN_spin}) and (\ref{eq:Heff_model_NN_spin}) can be simulated using giant atoms in a setup as sketched in \figpanel{fig_4_atom_1}{b}. Let us consider the Liouvillian dynamics of such a simulator with 4 giant atoms:
\begin{align} \label{eq_4_atom_simulator}
H(t) &= \sum_{n,m} g_{nm}(\omega_n, \omega_{m}) \mleft( \sigma_n^+ \sigma_{m}^- + \text{H.c.} \mright) + \sum_n \Omega_n(t) \sigma_n^x , \\
\mathcal{L}(t)\rho &= - i \mleft[ H(t), \rho \mright] + \sum_n \Gamma_n(\omega_n) \mathcal{D}[\sigma_n^-]\rho \\
+ \sum_{n,m} &\: \Gamma_{nm}(\omega_n,\omega_m) \mleft[ \mleft( \sigma_n^- \rho \sigma_m^+ - \frac{1}{2} \mleft\{ \sigma_n^+ \sigma_m^-, \rho \mright\} \mright) + \text{H.c.} \mright]. \nonumber
\end{align}
Here, we are in a frame rotating with the qubit frequencies, which we assume to all be $\omega$. For that case, the dependence of the couplings $g_{nm}$ and the decay rates $\Gamma_n$ and $\Gamma_{nm}$ on $\omega$ are shown in \figpanels{fig_4_atom_1}{c}{d}, where $\omega_0 = 2 \pi v / (\Delta x_1 + 2 \Delta x_2)$. Due to the identical spacing of the coupling points of each qubit, all the $\Gamma_n$ are equal. Additionally, we have $g_{12} = g_{23} = g_{34}$ and $g_{13} = g_{24}$, and the same equalities hold for $\Gamma_{nm}$.

From \figpanels{fig_4_atom_1}{c}{d}, we see that at the decoherence-free frequency $\omega_\text{DF}=2.5\omega_0$, all parameters are zero except for $g_{n, n+1}$ ($n = 1, 2, 3$). This allows us to perform two-qubit XY gates on all neighboring qubits simultaneously. Additionally, from \secref{sec3.1} we know that for two qubits far detuned from each other, the coupling between them is effectively 0. Thus, we can select to only perform some XY gates between some nearest neighbors. For example, by setting $\omega_{1, 2, 4} = \omega_\text{DF}$ and $\omega_3 = 3.5 \omega_0$ (such that $\Gamma_3 = 0$), we have $g_{12} \neq 0$ while $g_{23} = g_{34} = 0$. This allows us to perform a two-qubit XY-gate (and thus universal two-qubit operations, when adding single-qubit gates) on only qubits 1 and 2. Two-qubit XY gates on other neighboring qubits can be performed selectively in the same manner. 

To simulate the dynamics of single-qubit decay, we just have to let all neighboring qubits have different frequencies such that $g_{n,n+1} = 0$, and let the specific qubit have a frequency such that it decays. As we thus can perform both universal gates on neighboring qubits and selectively turn on and off single-qubit decay, this setup allows us to simulate \eqref{eq:L_model_NN_spin} in different parameter regimes by Trotterization as demonstrated in \secref{sec3.1}.

We note that a small-atom quantum simulator with parametric couplers between neighboring qubits and between qubits and the waveguide, as sketched in \figpanel{fig_4_atom_1}{e}, would also be able to simulate the model \eqref{eq:L_model_NN_spin}. However, such a setup for an $N$-spin model requires $2N - 1$ parametric couplers, which should be compared with zero for our giant-atom quantum simulator. These parametric couplers usually consist of qubits~\cite{McKay2016, Roth2017, Bengtsson2020, Ganzhorn2020, Kosen2022, Warren2023}. Thus, compared to a small-atom quantum simulator, the giant-atom quantum simulator requires fewer hardware resources, even when taking into account that a small-atom quantum simulator could work with fixed-frequency qubits, which do not require a flux line for their control.

The giant-atom quantum simulator in \figpanel{fig_4_atom_1}{b} can be readily realized with superconducting circuits as sketched in \figpanel{fig_4_atom_1}{f}, where a bent waveguide allows each qubit to couple to it at multiple points. The flux lines and resonators coupled to the qubits enable tuning the qubit frequency and applying a drive to or read out the qubit state, respectively. Note that this architecture can be realized on a single two-dimensional chip; there is no need for a three-dimensional flip-chip architecture to fit and address all components. 

With the physical parameters the same as those considered in \secref{sec3.1}, we have $\Delta x_1 + \Delta x_2 \approx \qty{6.77}{\cm}$ in \figpanel{fig_4_atom_1}{b}. With state-of-the-art 
techniques, a waveguide on a chip can be made at least $\qty{68}{\cm}$ long~\cite{Sundaresan2015}, and this allows a simulator with 10 giant atoms. We note, however, that this constraint is mainly due to $\omega_0$ should not be large, to prevent extra leakage of the qubit into other environments. Since $\omega_0$ is proportional to the speed of light in the waveguide $v$, this constraint can be softened by lowering $v$~\cite{Kuzmin2019}.


\subsection{Simulation of driven-dissipative spin chains with long-range interactions}
\label{sec_long_range}

\begin{figure*}
\center
\includegraphics[width=\linewidth]{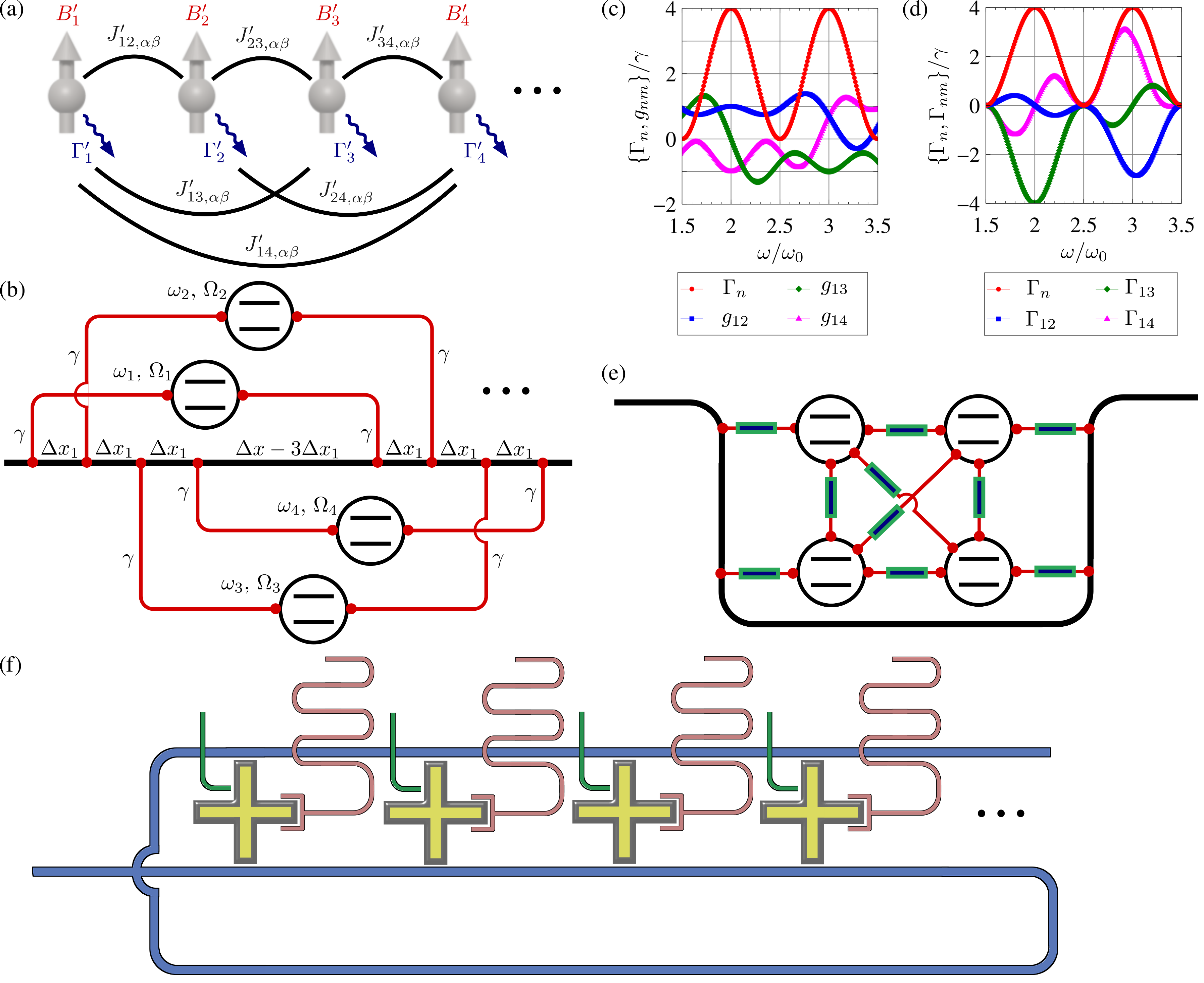}
\caption{A giant-atom quantum simulator for the simulation of driven-dissipative spin chains with long-range interactions.
(a) A sketch of the dissipative spin chain model [\eqref{eq:L_model_long_spin}] to be simulated.
(b) A setup for the giant-atom quantum simulator, where all pairs of qubits are coupled to the waveguide in a braided configuration.
(c,d) Qubit-qubit couplings $g_{nm}$ and decay rates $\Gamma_n$ and $\Gamma_{nm}$ as functions of qubit frequencies $\omega$ for the simulator in panel (b) with four giant atoms and $\Delta x = 8 \Delta x_1$.
(e) A small-atom quantum simulator to simulate the model in panel (a), where $N(N+1)/2$ parametric couplers (blue rectangles with green edges) arranged in a complex configuration are needed to tune the parameters in panels (c) and (d) for an $N$-qubit simulator.
(f) Sktech of a possible experimental setup for the giant-atom quantum simulator with superconducting qubits (yellow) coupled to the waveguide (blue). Flux lines (green) are used to tune the qubit frequencies and resonators (pink) are coupled to the qubits to enable single-qubit drives and measurements.}
\label{fig_4_atom_2}
\end{figure*}

We now consider the simulation of a driven-dissipative spin chain with long-range interactions, as illustrated in \figpanel{fig_4_atom_2}{a}. The relevant equations to simulate for this system are, in the frame rotating at the frequency of the spins,
\begin{align}
\tilde{H} &= \sum_{n, m, \alpha, \beta} \tilde{J}_{nm, \alpha\beta} S_n^\alpha S_m^\beta + \sum_n \tilde{B}_n S^x_n , \label{eq:H_model_long_spin} \\
\tilde{\mathcal{L}}\rho &= - i \mleft[ \tilde{H}, \rho \mright] + \sum_n \tilde{\Gamma}_n \mathcal{D}[\sigma_n^-]\rho , \label{eq:L_model_long_spin} \\
\tilde{H}_\text{eff} &= \tilde{H} - i \sum_n \frac{\Gamma'_n}{4} \sigma^z_n , \label{eq:Heff_model_long_spin}
\end{align}
where the notation is the same as in Eqs.~(\ref{eq:H_model_NN_spin})--(\ref{eq:Heff_model_NN_spin}).
This general model includes a wide range of dissipative spin models of recent interest~\cite{Catalano2023, Ricottone2020, Teretenkov2024}. Moreover, since spin systems are related to interacting fermions in one dimension through the Jordan--Wigner transformation~\cite{Jordan1928}, and in two dimensions through the Schrieffer--Wolff transformation~\cite{BRAVYI20112793}, being able to simulate this model would also enable investigations of the effects of many-body interactions in dissipative fermionic systems with long-range hoppings~\cite{rafiulislam2023twisted, Wang2023}. 

We put forward the giant-atom quantum simulator sketched in \figpanel{fig_4_atom_2}{b} for the quantum simulation of the dynamics in Eqs.~(\ref{eq:L_model_long_spin}) and (\ref{eq:Heff_model_long_spin}). Unlike the setup in \figpanel{fig_4_atom_1}{b}, where only neighboring qubits are coupled to the waveguide in a braided configuration, the arrangement of coupling points in \figpanel{fig_4_atom_2}{b} is such that all qubits are coupled to the waveguide in a braided configuration. This arrangement thus essentially allows decoherence-free coupling between all pairs of qubits at the decoherence-free frequency $\omega_\text{DF}$, which in turn enables the simulation of long-range spin interactions. 

We illustrate the all-to-all connectivity by considering the setup with four giant atoms in \figpanel{fig_4_atom_2}{b}. The Liouvillian dynamics of this simulator is given by the same master equation [\eqref{eq_4_atom_simulator}] as in the preceding subsection, where the parameters now have a different frequency dependence, as shown in \figpanels{fig_4_atom_2}{c}{d}, with $\omega_0 = 2 \pi v / \Delta x$. Due to the identical spacings between coupling points of each qubit, all the $\Gamma_n$ are equal. Additionally, we have $g_{12} = g_{23} = g_{34}$ and $g_{13} = g_{24}$ due to the symmetry of the setup, and the same equalities also hold for $\Gamma_{nm}$.

We see that, unlike in the setup for nearest-neighbor interactions in \figpanel{fig_4_atom_1}{b}, the long-range qubit-qubit couplings $g_{13}$ and $g_{24}$ are non-zero at the decoherence-free frequency $\omega_\text{DF} = 2.5 \omega_0$ in \figpanels{fig_4_atom_2}{c}{d}. These decoherence-free couplings allow us to perform long-range XY gates on pairs of distant qubits. For example, setting $\omega_{1,4} = \omega_\text{DF}$, $\omega_2 = 3.5 \omega_0$, and $\omega_3 = 1.5\omega_0$, the only non-zero parameter in $\mathcal{L}(t)$ is $g_{14}$. This enables the execution of a two-qubit XY gate on qubits 1 and 4. Single-qubit decays can be simulated in a similar manner as with the setup in \figpanel{fig_4_atom_1}{b}; see \secref{sec:nearest-neighbor}.

When comparing this giant-atom quantum simulator with other setups using small atoms, we note that performing long-range two-qubit gates in a small-atom quantum simulator represents a technical challenge. Even though a small-atom quantum simulator with four qubits arranged as in \figpanel{fig_4_atom_2}{e} allows to simulate \eqref{eq:L_model_long_spin} with four spins, it faces two challenges when scaling up. The first is the number of parametric couplers. To simulate \eqref{eq:L_model_long_spin} with $N$ spins, $N(N+1)/2$ parametric couplers are needed. This quadratic scaling results in a large cost in the physical setup. Additionally, the complexity of the setup increases with $N$ since the $N(N+1)/2$ parametric couplers need to be isolated from each other, which requires complex chip design. Finally, this setup is also limited by the number of parametric couplers that can be coupled to a single qubit. Thus, it appears much easier to achieve all-to-all coupling in a giant-atom quantum simulator than in a small-atom one.

Furthermore, compared to conventional setups to achieve all-to-all coupling, e.g., all qubits dispersively coupled to one and the same resonator~\cite{DiCarlo2010}, the giant-atom quantum simulator has two advantages. First, the couplings in a giant-atom quantum simulator are tunable. Second, a major challenge in conventional setups is the unwanted coupling between qubits when their detuning is small, which becomes inevitable when more qubits are added as there is a frequency range where the qubits work. The giant-atom quantum simulator can address this problem by reducing $\omega_0$, which can be done by either reducing the speed of light $v$ or increasing the waveguide length $\Delta x$.  

Finally, the giant-atom quantum simulator in \figpanel{fig_4_atom_2}{b} can be readily realized with superconducting circuits as sketched in \figpanel{fig_4_atom_2}{f}. Here, the resonators and flux lines can go over the waveguide without crossing interrupting it by using air bridges~\cite{Dunsworth2018} or multi-layer chips~\cite{Rosenberg2017, Kosen_2022, kosen2024signal}. This structure is scalable not only because of only needing $N$ qubits to simulate an $N$-spin system, but also because the structure complexity does not increase with $N$. New qubits can simply be added at the end of the qubit chain, which is much simpler than extending the small-atom quantum simulator in \figpanel{fig_4_atom_2}{e}.

There are two main limitations for the number of qubits $N$ that this implementation of a giant-atom quantum simulator may face: (i) the physical length of the waveguide, which is approximately $3 N \Delta x_1$, and (ii) the magnitude of $\omega_0$, which should be much larger than $g_{nm}$ such that the effective coupling between two detuned qubits is negligible. To have sufficient spacing between qubits such that a resonator can fit in in \figpanel{fig_4_atom_2}{f}, we assume $\Delta x_1 = \qty{1}{mm}$. Thus, a waveguide of length $\qty{68}{\cm}$, which has been demonstrated in experiment~\cite{Sundaresan2015}, allows the giant-atom quantum simulator to have more than 200 qubits in this configuration. Interestingly, this shows that the setup here is more compact than the setup considered for nearest-neighbor interactions in \secref{sec:nearest-neighbor}. To fulfill constraint (ii), note that $g_{nm}$ is of the same magnitude as $\gamma$. Thus $\omega_0 = 2 \pi v / \Delta x \ll g_{nm} \approx \gamma$ implies $\Delta x \ll 2 \pi v / \gamma$, which for $\gamma / (2\pi) = \qty{1}{\mega\hertz}$ gives $\Delta x \ll \qty{130}{m}$; this is clearly fulfilled even for several hundred qubits.


\subsection{Potential limitations}
\label{sec:limitations}

We now discuss potential limitations for scaling up our simulation protocol to larger systems, beyond what we already mentioned at the end of \secref{sec:nearest-neighbor} and \secref{sec_long_range}. The first limitation to consider is non-Markovian effects, which become non-negligible when the time $\tau = \Delta x / v$ it takes to travel between two coupling points relevant for the dynamics no longer satisfies $\gamma \tau \ll 1$, where $\Delta x$ is the distance between the coupling points and $v$ is the speed of light in the waveguide. When more qubits are added to the simulation, $\Delta x$ inevitably increases, and non-Markovian effects will eventually begin to play an important role. In this manuscript, we have mainly considered typical, but conservative, parameter values of $\gamma / (2\pi) = \qty{1}{\mega\hertz}$ and $v = \qty{1.3e8}{\meter/\second}$; these values yield $\Delta x \ll \qty{20}{\meter}$. Thus non-Markovian effects are not expected to play an important role for the scaled-up version of the giant-atom-based simulators until we reach several tens or several hundreds of qubits, depending on the setup.

We note that for a larger $\gamma$ or a smaller $v$, such as with surface acoustic waves~\cite{Andersson2019, Guo2017}, and structured environments~\cite{Calajo2016}, non-Markovian effects can occur for shorter distances between coupling points. Additionally, with $\qty{30}{m}$ long waveguides realized in recent experiments~\cite{Storz2023}, non-Markovian effects can also take place. A quantitative analysis of non-Markovian effects in the scaled-up version of the giant-atom quantum simulator remains an open challenge~\cite{Guo2017, Guo2020}, and is left for future work. Importantly, we note that this challenge can also be an opportunity for realizing quantum simulations of non-Markovian systems, which we also plan to address in future work.

Another challenge faced when the system size increases is that when switching between different simulation regimes in the Trotter steps, an unwanted decay on the qubits can appear. For example, consider performing an XY gate on qubits 1 and 3 after an XY gate on qubits 1 and 4 in the giant-atom quantum simulator in \figpanel{fig_4_atom_2}{b}. This requires tuning the frequencies from $\{ \omega_1, \omega_2, \omega_3, \omega_4 \} = \{ \omega_\text{DF}, 3.5 \omega_0, 1.5 \omega_0, \omega_\text{DF} \}$ to $\{ \omega_1, \omega_2, \omega_3, \omega_4 \} = \{ \omega_\text{DF}, 3.5 \omega_0, \omega_\text{DF}, 1.5 \omega_0 \}$. During this process, both qubits 3 and 4 will be tuned to through frequencies where they decay. To reduce this effect, we can increase the speed of tuning the frequency $v_1$ or reduce $\omega_0$. With the distance $\Delta x = \qty{68}{\cm}$ between coupling points that can be realized with the state-of-the-art techniques, $\omega_0 / (2\pi) \approx \qty{0.19}{\giga\hertz}$, and can be further reduced by reducing the speed of light $v$. On the other hand, $v_1 / (2\pi)$ has a typical value of $0.1 \sim \qty{1}{\giga\hertz} / \qty{}{\nano\second}$~\cite{Collodo2020}. Thus, the time for tuning the frequency can be lowered to around $\qty{1}{\nano\second}$ to reduce the effect of the unwanted decay, which is much smaller than the typical simulation time of $\sim \qty{1}{\micro\second}$ we considered in our examples. 


\section{Conclusion}
\label{Ch_conclusion}

We have introduced giant atoms as a new paradigm for quantum simulation of generic open quantum many-body systems. The giant-atom-based quantum simulators we propose are simultaneously scalable and highly tunable, distinguishing them from other proposals and implementations that generally offer only one of these benefits. 

After first outlining the general idea of how to use giant atoms for quantum simulation, we studied an example of quantum simulation in great detail to make the idea more concrete. In the example, we showed how a giant-atom quantum simulator using two giant atoms can simulate a qubit coupled to a driven-dissipative qubit. In particular, we showed how different parameter regimes for this open quantum system can be simulated by only controlling the frequency of one giant atom. This simulation enabled us to characterize the quantum Zeno crossover in the Liouvillian dynamics and the transition from oscillatory to non-oscillatory dynamics in the effective non-Hermitian Hamiltonian dynamics of the two-qubit system. This demonstration highlighted the high tunability of giant atoms. 

We analyzed the robustness of the two-qubit simulation results against extra decay and dephasing in noisy qubits, and discussed other possible experimental imperfections, showing that it is realistic to implement this quantum simulation with good accuracy in existing experimental systems. Finally, we presented how the giant-atom quantum simulator can be scaled up to simulate generic dissipative spin systems, including ones with long-range couplings, demonstrating its advantages over conventional small-atom quantum simulators in terms of the number of components needed. We also provided concrete calculations of relevant parameters for experimental realizations of the scaled-up simulators with superconducting qubits, and discussed potential limitations to further scaling up the simulators. 

We note that, recently, much effort has gone into improving simulations of open quantum many-body systems by optimizing the simulation algorithm~\cite{Kamakari2022, Guimaraes2023, peetz2023simulation, Leppakangas2023}. Our work, on the other hand, focuses on advancing quantum simulation by a new physical setup to simulate open quantum many-body systems, and may be combined with these new algorithms for more efficient simulations of open quantum many-body systems.

We also note that, while we focus on superconducting qubits for the physical realization of our giant-atom quantum simulator, it can also be realized on other platforms such as microwave cavities or spin ensembles~\cite{Karg2019, Wang2022}, where the multi-level nature of the cavities or ensembles may enable the simulation of dissipative spin models with spins larger than $1/2$. Another interesting possible physical realization of our simulation scheme would be the implementation of giant atoms proposed with cold atoms in an optical lattice~\cite{Gonzalez-Tudela2019}.

Extending the analysis of the giant-atom quantum simulator to giant atoms with more levels in various configurations is one potential research direction. As discussed in \secref{sec:limitations}, another possible extension of the scheme is to non-Markovian dynamics, which could be realized by increasing the distance between coupling points of the giant atoms, or by reducing the speed of light in the waveguide. Finally, since the analysis of the giant-atom quantum simulator here was quite general, a more detailed analysis for some specific implementations of models to simulate would be desirable, to determine which models would be most suitable for first experiments at a larger scale.


\begin{acknowledgments}
    
We acknowledge fruitful discussions with Liangyu Chen, Fei Song, Bharath Khannan, Aziza Almanakly, Beatriz Yankelevich, Laura Garc\'{i}a-\'{A}lvarez, Lei Du, and Ariadna Soro \'{A}lvarez. The numerical calculations were performed using the QuTiP library~\cite{Johansson2012, Johansson2013}. We acknowledge support from the Swedish Foundation for Strategic Research (grant number FFL21-0279). AFK is also supported by the Swedish Research Council (grant number 2019-03696), the Swedish Foundation for Strategic Research (grant number FUS21-0063), the Horizon Europe programme HORIZON-CL4-2022-QUANTUM-01-SGA via the project 101113946 OpenSuperQPlus100, and the Knut and Alice Wallenberg Foundation through the Wallenberg Centre for Quantum Technology (WACQT).

\end{acknowledgments}


\appendix


\section{Dynamics of the model with a qubit coupled to a driven-dissipative qubit}
\label{app_L}

Here we provide some further details about the model with one qubit coupled to a driven-dissipative qubit, which was introduced in \secref{Ch_model} and used as a prototype model to simulate with our giant-atom quantum simulator in \secref{Ch_sim}. We discuss the dynamics of the master equation for this model [\eqref{eq_model_L} in \secref{Ch_model}] with the initial state $\rho(0)=\mleft( \ket{0}_1 \otimes \ket{1}_2 \mright) \mleft( \bra{0}_1 \otimes \bra{1}_2 \mright)$.

Let $\{\omega_n\}$ be the eigenvalues of $\mathcal{L}'$ from \eqref{eq_model_L} with $\{\rho_{n,R(L)}\}$ the corresponding right (left) eigen-density matrices. Expanding $\rho(0)$ in the eigenbasis $\{\rho_{n,R}\}$ as $\rho(0) = \sum_n c_n \rho_{n,R}$, where $c_n = \text{Tr}\mleft[ \rho_{n,L}^\dag \rho(0) \mright]$, we have
\begin{align}
\rho(t') &= \exp \mleft( \mathcal{L}' t' \mright) \rho(0) \nonumber \\
&= \sum_n c_n \exp \mleft( \mathcal{L}' t' \mright) \rho_n \nonumber \\
&= \sum_n c_n \exp \mleft( \omega_n t' \mright) \rho_n.
\label{eqA1}
\end{align}
Since $\mathcal{L}'$ is completely positive and trace-preserving, all its eigenvalues have real parts $\Re$ less than or equal to zero. Furthermore, since $\text{Tr} \mleft[ \rho(t') \mright] \equiv 1$, $\mathcal{L}'$ must have at least one eigenvalue equal to 0; the corresponding right density matrix $\rho_\text{ss}$ is the steady state with $\text{Tr} \mleft[ \rho_\text{ss} \mright] = 1$. For the model in \eqref{eq_model_L}, $\mathcal{L}'$ has a unique steady state. 

\begin{figure*}
\center
\includegraphics[width=\linewidth]{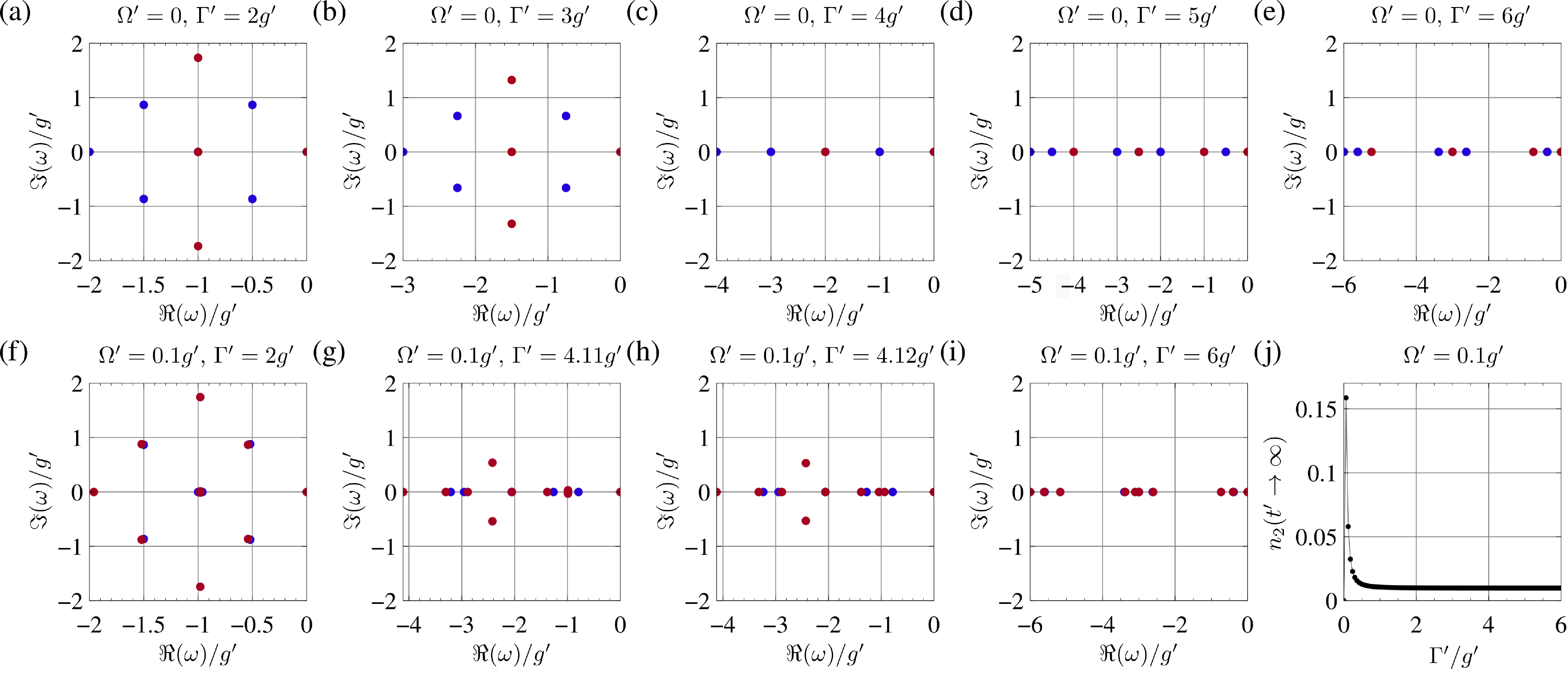}
\caption{Spectrum and $n_2(\tilde{t}\to\infty)$ for the Liouvillian in \eqref{eq_model_L}.
(a-e) Liouvillian spectrum with $\Omega' = 0$. The eigenvalues whose corresponding eigen-density matrices have overlap with $\rho(0) = \ket{0}_1 \otimes \ket{1}_2 \bra{0}_1 \otimes \bra{1}_2$ are highlighted with dark red color. The spectrum reveals the quantum Zeno crossover at $\Gamma' = 4 g'$.
(f-i) Liouvillian spectrum with $\Omega' = 0.1g'$. The spectrum reveals the quantum Zeno crossover at around $\Gamma' = 4.11g'$.
(j) $n_2(\tilde{t}\to\infty)$ for the Liouvillian in \eqref{eq_model_L} with $\Omega' = 0.1 g'$ as a function of $\Gamma'$.}
\label{figA1}
\end{figure*}

We can thus write \eqref{eqA1} as
\begin{align}
\rho(t') &= \rho_\text{ss} + \sum_{n \neq \text{ss}} c_n \exp \mleft( \omega_n t' \mright) \rho_n \nonumber \\
&\approx \rho_\text{ss} + \sum_m c_m \exp \mleft( \omega_m t' \mright) \rho_m \quad (t' \to \infty) ,
\end{align}
where $\{\omega_m\}$ are the nonzero eigenvalues of $\mathcal{L}'$ with the largest real part, $\rho_m$ are the corresponding eigen-density matrices, and $c_m \neq 0$ are the overlaps between $\rho(0)$ and $\rho_m$. We see that the long-time behavior of $\rho(t')$ is determined by $\Re(\omega_m)$.

In \figpanels{figA1}{a}{e}, we plot the Liouvillian spectrum of \eqref{eq_model_L} for $\Omega'=0$ at different $\Gamma'$. The eigenvalues whose corresponding eigen-density matrices overlap with $\rho(0)$ are highlighted with dark red color. By extracting $\Re(\omega_m)$ in all these cases, we obtain the long-time behavior of the effective relaxation rate in \figpanel{fig_model}{d}. We note that the quantum Zeno crossover at $\Gamma' = 4 g'$ is related to the parity-time (PT) transition in the Liouvillian spectrum~\cite{Minganti2019}, where all the eigenvalues of $\mathcal{L}'$ become real.

Similarly, for $\Omega'=0.1g'$, the Liouvillian spectrum is shown in \figpanels{figA1}{f}{i}. Compared to the case $\Omega' = 0$, here $\rho(0)$ has overlap with states having larger real eigenvalues, resulting in slower long-time decay, as shown in \figpanel{fig_model}{g}. Additionally, the quantum Zeno crossover point becomes shifted, and is only related to the PT transition of two specific eigenvalues [see \figpanels{figA1}{g}{h}].

As a final point, we discuss how the quantum Zeno crossover in both cases can be revealed through the behavior of the population in qubit 2, $n_2(\tilde{t}) = \mleft\{ 1 + \text{Tr} \mleft[ \sigma_2^z \rho(\tilde{t}) \mright] \mright\} / 2$. Since $n_2(\tilde{t}\to\infty) = \mleft (1 + \text{Tr} \mleft[ \sigma_2^z \rho_\text{ss} \mright] \mright) / 2$, we have
\begin{align}
& n_2(\tilde{t}) - n_2(\tilde{t} \to \infty) = \nonumber\\
& \frac{1}{2} \sum_m c_m \exp \mleft( \omega_m t' \mright) \text{Tr} \mleft[ \sigma_2^z \rho_m \mright] \quad (t' \to \infty) ,
\end{align}
and therefore this quantity reveals the relaxation rate due to $\Re(\omega_m)$. For $\Omega'=0$, the steady state of $\mathcal{L}'$ is $\rho_\text{ss} = \ket{0}_1 \otimes \ket{0}_2 \bra{0}_1 \otimes \bra{0}_2$ with $n_2(\tilde{t} \to \infty) = 0$; for $\Omega'=0.1g$, the steady state has a finite nonzero population $n_2(\tilde{t} \to \infty) \neq 0$ [see \figpanel{figA1}{j}].


\section{Post-selection}
\label{app_H}

Here we briefly review how post-selection works and show how it results in the effective non-Hermitian Hamiltonian dynamics in Eqs.~(\ref{eq_model_Heff}) and (\ref{eq_sim_Heff}). Under the Markovian approximation, the time evolution of the system density matrix, $\rho(t) = F_t \rho(t = 0) = \exp \mleft( \mathcal{L} t \mright) \rho(t=0)$, only depends on the infinitesimal evolution $F_{dt}$:
\beq
\rho(t + dt) = F_{dt} \rho(t) = \mleft[ \mathbf{I} + \mathcal{L} dt + O(dt^2) \mright] \rho(t) ,
\eeq
where $\mathbf{I}$ is the identity matrix. From the Choi--Kraus theorem~\cite{CHOI1975285, nielsen_chuang_2010, Preskill_note} we know that the above evolution also can be represented as Kraus operators:
\beq
F_{dt} \rho(t) = \sum_\alpha K_{\alpha, dt} \rho(t) K_{\alpha, dt}^\dag
\eeq
with $\sum_\alpha K_{\alpha, dt}^\dag K_{\alpha, dt} = \mathbf{I}$.

In particular, with
\begin{align}
K_{0, dt} &= \mathbf{I} - i \mleft[ \tilde{g} \mleft( \sigma_1^+ \sigma_2^- + \text{H.c.} \mright) + \tilde{\Omega} \sigma_1^x - i \tilde{\Gamma} \mleft( \sigma^z_1 + \mathbf{I} \mright) / 4 \mright] dt , \\
K_{1, dt} &= \sqrt{\tilde{\Gamma}} \sigma^-_1 \sqrt{dt}
\end{align}
we obtain the dynamics given by \eqref{eq_model_L}. The time evolution governed by $K_{1,dt}$ describes a jump of the qubit from its excited state to its ground state with a photon emitted to the environment (in this case, the waveguide). Thus, if no photons are observed in the environment during the time interval $dt$ in an experiment, the system is known to have undergone the evolution governed by $K_{0, dt}$. 

By successively measuring the environment and selecting results where no photon has been observed in the environment during any small time interval $dt$, the selected results thus follow the dynamics governed by $K_{0, dt}$:
\begin{align}
\rho(t+dt) &= K_{0, dt} \rho(t) K_{0, dt}^\dag \nonumber \\
&= \rho(t) - i \mleft[ H_\text{eff} \rho(t) - \rho(t) H_\text{eff}^\dag \mright] + O(dt^2) ,
\end{align}
with the effective non-Hermitian Hamiltonian $H_\text{eff}$ in \eqref{eq_model_Heff}. This yields the time evolution
\beq
\rho(t) = \exp \mleft( - i H_\text{eff} t \mright) \rho(0) \exp \mleft( i H_\text{eff}^\dag t \mright) ,
\label{eq:app_post-selection_rhot}
\eeq
where the norm $|\rho(t)|$ describes the probability of having the selected dynamics in all experimental results, and the normalized density matrix $\rho(t) / |\rho(t)|$ is the state after these selected dynamics.

The derivation of the effective non-Hermitian Hamiltonian dynamics for \eqref{eq_simulator_L} is similar. The dynamics given by \eqref{eq_simulator_L} can be written as~\cite{Kockum2018}
\begin{align}
&\mathcal{L}(t)\rho = -i \mleft[ g(\omega_1, \omega_2) \mleft( \sigma_1^+ \sigma_2^- + \text{H.c.} \mright) + \Omega_1(t) \sigma_1^x + \Omega_2(t) \sigma_2^x, \rho \mright] \nonumber \\
&+ \mathcal{D} \mleft[ \mleft( e^{i (\varphi_0 + \varphi_1)} + e^{i \varphi_1} \mright) \sqrt{\frac{\gamma}{2}} \sigma^-_1 + \mleft( e^{i \varphi_0} + 1 \mright) \sqrt{\frac{\gamma}{2}} \sigma^-_2 \mright] \rho \nonumber \\
&+ \mathcal{D} \mleft[ \mleft( e^{i \varphi_0} + 1 \mright) \sqrt{\frac{\gamma}{2}} \sigma^-_1 + \mleft( e^{i (\varphi_0 + \varphi_1)} + e^{i \varphi_1} \mright) \sqrt{\frac{\gamma}{2}} \sigma^-_2 \mright] \rho , 
\end{align}
where $\varphi_0 = 2 \pi \omega / \omega_0$ and $\varphi_1 = \varphi_0 \Delta x_1 / (\Delta x_1 + \Delta x_2)$. The time evolution can be represented using the Kraus operators
\begin{align}
K_{0, dt} &= \mathbf{I} - i H_\text{eff} dt , \\
K_{1, dt} &= \mleft[ \mleft( e^{i (\varphi_0 + \varphi_1)} + e^{i \varphi_1} \mright) \sqrt{\frac{\gamma}{2}} \sigma^-_1 \mright. \nonumber \\
&\quad \mleft. + \mleft( e^{i \varphi_0} + 1 \mright) \sqrt{\frac{\gamma}{2}} \sigma^-_2 \mright] \sqrt{dt} , \\
K_{2, dt} &= \mleft[ \mleft( e^{i \varphi_0} + 1 \mright) \sqrt{\frac{\gamma}{2}} \sigma^-_1 \mright. \nonumber \\
&\quad \mleft. + \mleft( e^{i (\varphi_0 + \varphi_1)} + e^{i \varphi_1} \mright) \sqrt{\frac{\gamma}{2}} \sigma^-_2 \mright] \sqrt{dt} ,
\end{align}
where $H_\text{eff}$ is given by \eqref{eq_sim_Heff}. Thus, when quantum jumps do not occur in the system, its dynamics are given by the effective non-Hermitian Hamiltonian $H_\text{eff}$ in \eqref{eq_sim_Heff}.


\section{Protocol to tune giant-atom frequency}
\label{app_omega}

In \eqref{eq_omega}, we presented the way to tune the giant-atom frequency $\omega_1(t)$ in the giant-atom quantum simulator for the two-qubit model in \secref{Ch_model}. There, we noted that $t_1$ and $t_2$ are determined by
\beq
\int_0^{t_1 + t_2} \Gamma_1[\omega_1(t)] dt = \tilde{\Gamma} \tilde{t} / l. 
\eeq
We here present the exact formulas for $t_1$ and $t_2$, i.e., the time spent tuning the qubit's frequency and the time that the qubit remains at its maximum decay rate, respectively. 

The first thing to note is that $\Gamma_1[\omega_1(t)]$ reaches its maximum value $\Gamma_\text{max}$ at $\omega_1 \pm 0.5 \omega_0$. If $\int_0^{t_1} \Gamma_1[\omega_1(t)] dt = \tilde{\Gamma} \tilde{t} / l$ is already satisfied before $\Gamma_1$ reaches $\Gamma_\text{max}$, we know that $t_2 = 0$. In particular, since we are tuning the frequency at a speed $v_1$, the time it takes to reach $\Gamma_\text{max}$ from $\omega_1 = \omega_\text{DF}$ is $0.5 \omega_0 / v_1$, and the total time spent in tuning $\omega_1$ is $2 \omega_0 / v_1$. This yields 
\begin{align}
& \quad \int_0^{2 \omega_0 / v_1} \Gamma_1[\omega_1(t)] dt \nonumber \\
&= 4 \int_0^{0.5 \omega_0 / v_1} \Gamma_1(\omega_\text{DF} + v_1 t) dt \nonumber \\
&= 8 \gamma \int_0^{0.5 \omega_0 / v_1} \mleft[ 1 + \cos \mleft( 5 \pi + 2 \pi v_1 t / \omega_0 \mright) \mright] dt \nonumber \\
&= 4 \gamma \omega_0 / v_1.
\end{align}

Thus, if $\tilde{\Gamma} \tilde{t} / l < 4 \gamma \omega_0 / v_1$, we have $t_2 = 0$, and $t_1$ given by
\beq
8 \gamma \int_0^{t_1} \mleft[ 1 + \cos \mleft( 5 \pi + 2 \pi v_1 t / \omega_0 \mright) \mright] dt = \tilde{\Gamma} \tilde{t} / l ,
\eeq
which yields
\beq
8 \gamma \mleft[ t_1 - \frac{\omega_0}{2\pi v_1} \sin \mleft( 2 \pi v_1 t_1 / \omega_0 \mright) \mright] = \tilde{\Gamma} \tilde{t} / l ,
\eeq
which can be solved for $t_1$. On the other hand, if $\tilde{\Gamma} \tilde{t} / l > 4 \gamma \omega_0 / v_1$, then we have $t_1 = 0.5 \omega_0 / v_1$, and $t_2 = \mleft( \tilde{\Gamma} \tilde{t} / l - 4 \gamma \omega_0 / v_1 \mright) / (4 \gamma)$.


\section{Protocol-independent errors in the giant-atom quantum simulator} 
\label{sec_app_err}

In this appendix, we give further details about two types of potential simulation errors, beyond those analyzed in more detail in \secref{Ch_err}: statistical errors and imperfect post-selection. For the analysis, we stick to the illustrative example of simulating one qubit coupled to a driven-dissipative qubit, as described in Sections~\ref{Ch_model} and \ref{Ch_sim}. We note that both types of errors that we analyze here are not protocol-independent; they exist in general quantum simulators.


\subsection{Statistical errors} 
\label{sec_statistical_err}

To obtain a good estimate of an observable in a quantum simulation, generally a certain amount of repeated experiments have to be conducted. Here we discuss the potential statistical error resulting from an insufficient number of repeated experiments. We estimate the number of repeated experiments required to obtain a faithful simulation result for $n_2(t')$ (the population in qubit 2), and to correctly predict the quantum Zeno crossover point.

According to the central limit theorem~\cite{Fischer2011}, the error in $n_2(t')$ is smaller than $3\sqrt{n_2(t')(1-n_2(t'))/N_\text{exp}}$ in $N_\text{exp}$ measurements. Thus, the relative error in $n_2(t')$ with $N_\text{exp}$ measurements is
\beq
\delta_\text{exp}(t') = \frac{3 \sqrt{n_2(t') \mleft[ 1 - n_2(t') \mright]}}{\sqrt{N_\text{exp}} \mleft[ n_2(\tilde{t}) - n_2(\tilde{t} \to \infty) \mright]}.
\eeq
For concreteness, we take $\Gamma' = 6 g'$ as an example. For $\Omega' = 0$ at $t' = 3 \pi / g'$, the error is $\delta_\text{exp}(t') \approx 3 \sqrt{\qty{1e3}{} / N_\text{exp}}$. For $\Omega' = 0.1$ at $t' = 5 \pi / g'$, the error is $\delta_\text{exp}(t') \approx \qty{1e3}{} \sqrt{1 / N_\text{exp}}$. Having $\delta_\text{exp}(t') < 0.5$ would be sufficient; this value results in $N_\text{exp} = 4000$ for $\Omega' = 0$ at $t' = 3 \pi / g'$, and $N_\text{exp} = \qty{4e6}{}$ for $\Omega' = 0.1$ at $t' = 5 \pi / g'$. Note that, if $t'$ is decreased, the number of required experiments decreases exponentially. 

For the simulation of the effective non-Hermitian Hamiltonian dynamics in \eqref{eq_model_Heff}, we note that, as the simulation time increases, the probability of quantum jumps increases. Thus, to simulate the effective non-Hermitian Hamiltonian dynamics, more experiments have to be performed to have a sufficient amount of data remaining after post-selection. 

The probability of having no quantum jumps until $t'$ is $P(t') = \text{Tr}[\rho(t')]$, where $\rho(t')$ is given by \eqref{eq:app_post-selection_rhot} in \appref{app_H}. Thus, to have $n_\text{post}$ data points remaining after post-selection, $n_\text{post} / \text{Tr}[\rho(t')]$ experiments need to be performed. For observing the oscillation of the population of qubit 2, $n_\text{post} \sim 100$ would be sufficient. Since $\text{Tr}[\rho(t')]$ reaches its minimum of $\sim \qty{1e-4}{}$ at $t' = 3 \pi / g'$ for $\Gamma' = 3.9g'$, around $\qty{1e6}{}$ experiments have to be performed to faithfully simulate the effective non-Hermitian Hamiltonian dynamics. We note that a potential advantage of Trotter decomposition in this case is the ability to abort the experiment in the middle when a quantum jump occurs, which reduces the total simulation time.


\subsection{Imperfect post-selection}

\begin{figure}
\center
\includegraphics[width=\linewidth]{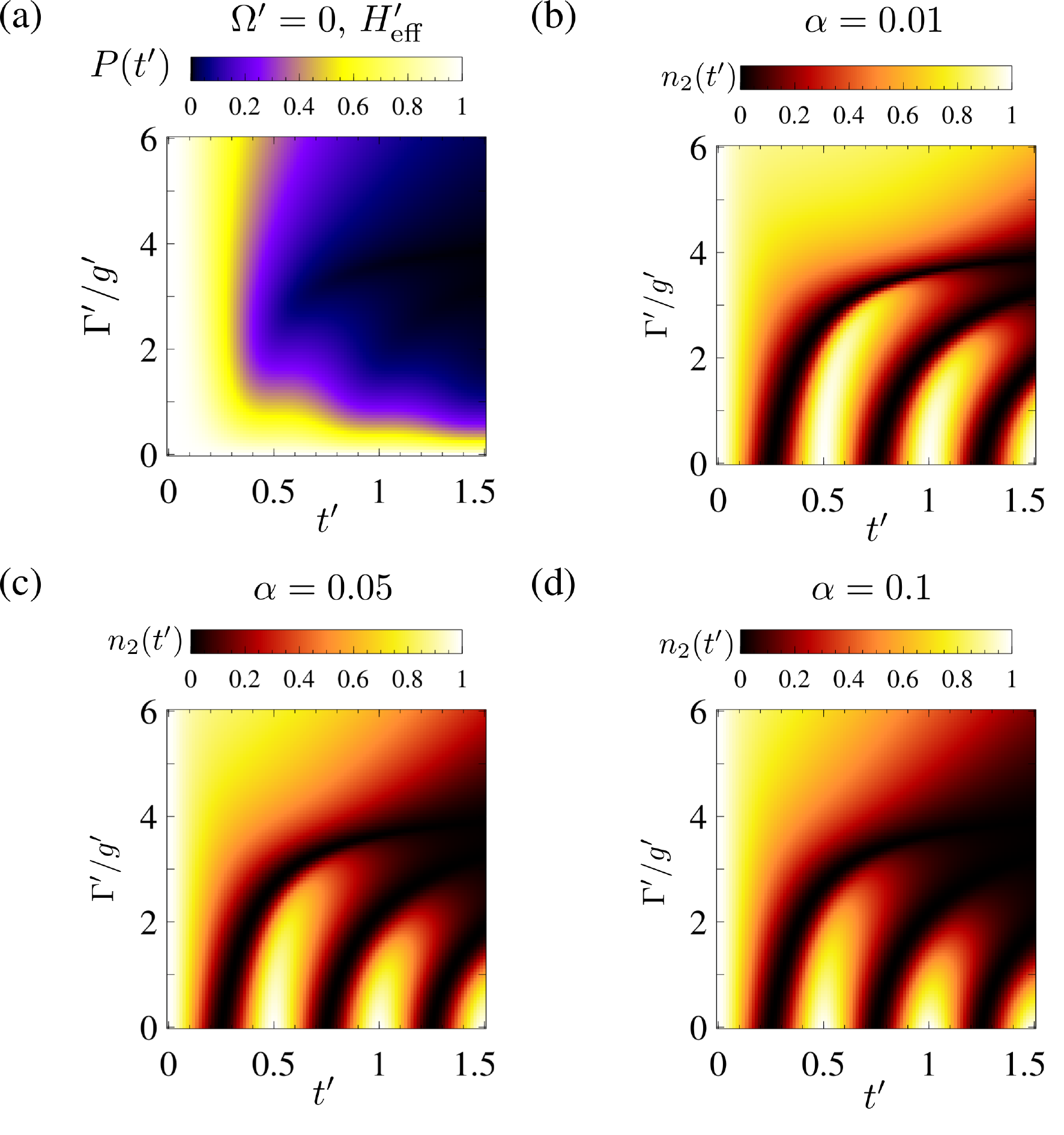}
\caption{Simulation error in the effective Hamiltonian dynamics due to imperfect post-selection.
(a) The probability of having the results with the effective Hamiltonian dynamics, given by $P(t')=\text{Tr}\rho(t')$.
(b-d) The simulated effective Hamiltonian dynamics under different photon detector error rates $\alpha$.}
\label{fig_bad_photon}
\end{figure}

We now analyze the influence of imperfect post-selection on the simulation of the effective non-Hermitian Hamiltonian dynamics. Such imperfections can be either due to a false quantum jump (dark count) or a false no-jump due to imperfect photon detectors.

When a false quantum jump occurs, a quantum jump has not actually taken place in the experiment, but the result is discarded due to the false jump. This will not change the simulated dynamics, but will result in more repeated experiments being needed to obtain a result with the same statistical certainty.

When a false no-jump occurs, the experimental result where a quantum jump has occurred is included in the simulated effective non-Hermitian Hamiltonian dynamics. Including this result changes the simulated dynamics. 

For the particular example of the two-qubit system we consider, when a quantum jump occurs, it always brings the system to $n_1 = n_2 = 0$. Thus, it will result in an additional decay of $n_2(t')$ compared to the effective non-Hermitian Hamiltonian dynamics. This additional decay will not change the transition from oscillatory to non-oscillatory dynamics, but will make it less visible.

To illustrate this effect, we consider a photon detector that reports false no-jumps with an error rate $\alpha$, i.e.., among all results where a quantum jump has occurred, $\alpha$ of them have been falsely reported as no-jump and are thus included in the simulated dynamics. Let $n_{2,\mathcal{L}}(t')$, $n_{2,H}(t')$, and $n_{2,j}(t')$ be the simulated population of qubit 2 under the Liouvillian dynamics, the effective non-Hermitian Hamiltonian dynamics, and the dynamics in which a quantum jump has occurred, respectively. We then have by definition that
\beq \label{eq_bad_photon_1}
n_{2, \mathcal{L}}(t') = P(t') n_{2, H}(t') + [1 - P(t')] n_{2, j}(t').
\eeq
The simulated population is thus
\beq \label{eq_bad_photon_2}
n_2(t') = \frac{P(t') n_{2, H}(t') + \alpha [1 - P(t')] n_{2,j}(t')}{P(t') + \alpha [1 - P(t')]} ;
\eeq
out of the $1-P(t')$ instances where quantum jumps occur, $\alpha$ of them are included in the simulated dynamics, which gives the factor of $\alpha [1 - P(t')]$ in $n_{2, j}(t')$.

Inserting \eqref{eq_bad_photon_1} into \eqref{eq_bad_photon_2}, we obtain
\beq \label{eq_n_bad}
n_{2}(t') = \frac{(1 - \alpha) P(t') n_{2, H}(t') + \alpha n_{2, \mathcal{L}}(t')}{P(t') + \alpha [1 - P(t')]}.
\eeq
As shown in \figpanel{fig_bad_photon}{a}, as $t'$ increases, $P(t')$ decreases, and thus the influence of the error in the photon detector on $n_2(t')$ is larger. For different values of $\alpha$, the simulated dynamics are shown in \figpanels{fig_bad_photon}{b}{d}. There we see that as $\alpha$ increases, more Liouvillian dynamics are involved in the simulation and the oscillation of $n_2(t')$ for small $\Gamma'$ becomes less visible. However, the value of $\Gamma'$ where the transition from oscillatory to non-oscillatory dynamics occurs is not influenced given sufficient accuracy of around $\qty{1e-3}{}$ of the simulated qubit population.



\bibliography{main, NH, GA}

\end{document}